\pdfoutput=1

\documentclass[11pt,twoside,a4paper,cmspaper,final,collab]{cms-tdr}
\def\svnVersion{325186}\def\cmsCernNo{2013-037}\def\cmsCernDate{\today}\def\cmsMessage{Published in the Journal of High Energy Physics as \href{http://dx.doi.org/10.1007/JHEP01(2016)006}{\doi{10.1007/JHEP01(2016)006.}}}

\begin{document}\cmsNoteHeader{HIN-14-010}

\hyphenation{had-ron-i-za-tion}
\hyphenation{cal-or-i-me-ter}
\hyphenation{de-vices}
\RCS$Revision: 313661 $
\RCS$HeadURL: svn+ssh://svn.cern.ch/reps/tdr2/papers/HIN-14-010/trunk/HIN-14-010.tex $
\RCS$Id: HIN-14-010.tex 313661 2015-12-07 00:59:57Z alverson $
\newcommand {\MPT}      {\ensuremath{\PTslash^{\parallel}}\xspace}
\newcommand {\MPTave}      {\ensuremath{\langle \PTslash^{\parallel}\rangle}\xspace}
\newcommand {\MPTpTdR}      {\ensuremath{\langle \PTslash^{\parallel} \rangle_{\pt^{\text{trk}},\Delta}}\xspace}
\newcommand {\MPTdR}      {\ensuremath{\langle \PTslash^{\parallel} \rangle_{\Delta}}\xspace}
\newcommand {\MPTpT}      {\ensuremath{\langle \PTslash^{\parallel} \rangle_{\pt^{\text{trk}}}}\xspace}
\newcommand {\MPTcum}      {\ensuremath{\langle \PTslash^{\parallel} \rangle_{[0,\Delta]}}\xspace}
\newcommand {\MPTsum}      {\ensuremath{\langle \PTslash^{\parallel} \rangle_{\Sigma}}\xspace}
\newcommand {\AJ}      {\ensuremath{A_J}\xspace}
\newcommand {\rootsNN}      {\ensuremath{\sqrt{s_{_\mathrm{NN}}}}\xspace}
	
\newlength\cmsFigWidth
\ifthenelse{\boolean{cms@external}}{\setlength\cmsFigWidth{0.85\columnwidth}}{\setlength\cmsFigWidth{0.4\textwidth}}
\ifthenelse{\boolean{cms@external}}{\providecommand{\cmsLeft}{top}}{\providecommand{\cmsLeft}{left}}
\ifthenelse{\boolean{cms@external}}{\providecommand{\cmsRight}{bottom}}{\providecommand{\cmsRight}{right}}
\ifthenelse{\boolean{cms@external}}{\providecommand{\cmsTableResize[1]}{\relax{#1}}}{\providecommand{\cmsTableResize[1]}{\resizebox{\columnwidth}{!}{#1}}}
\ifthenelse{\boolean{cms@external}}{\providecommand{\cmsTableResizeSmall[1]}{\relax{#1}}}{\providecommand{\cmsTableResizeSmall[1]}{\resizebox{0.7\columnwidth}{!}{#1}}}
\newcolumntype{x}{D{,}{.}{-1}}
\newcolumntype{y}[1]{D{,}{\, \pm \,}{#1}}

\cmsNoteHeader{HIN-14-010}
\title{Measurement of transverse momentum relative to dijet systems in PbPb and pp collisions at \texorpdfstring{$\rootsNN = 2.76\TeV$}{sqrt(s[NN]) = 2.76 TeV}}

\date{\today}

\abstract{
An analysis of dijet events in PbPb and pp collisions is performed to explore the properties of energy loss by partons traveling in a quark-gluon plasma. Data are collected at a nucleon-nucleon center-of-mass energy of 2.76\TeV at the LHC. The distribution of transverse momentum (\pt) surrounding dijet systems is measured by selecting charged particles in different ranges of \pt and at different angular cones of pseudorapidity and azimuth. The measurement is performed as a function of centrality of the PbPb collisions, the \pt asymmetry of the jets in the dijet pair, and the distance parameter R used in the anti-\kt jet clustering algorithm. In events with unbalanced dijets, PbPb collisions show an enhanced multiplicity in the hemisphere of the subleading jet, with the \pt imbalance compensated by an excess of low-\pt particles at large angles from the jet axes.
}

\hypersetup{%
pdfauthor={CMS Collaboration},%
pdftitle={Measurement of transverse momentum relative to the dijet system in PbPb and pp collisions at sqrt(sNN) = 2.76 TeV},%
pdfsubject={CMS},%
pdfkeywords={CMS, physics, software, computing}}

\maketitle

\section{Introduction}

Partons produced at large transverse momenta (\pt) through hard-scattering processes in heavy-ion
collisions are expected to lose energy as they travel through the quark-gluon plasma (QGP) created in these interactions~\cite{Bjorken:1982tu}.
Experiments at RHIC and the LHC have observed a suppression in the yield of
high-$\pt$ particles relative to suitably scaled pp collision data, and a significant reduction in back-to-back high-\pt hadron correlations \cite{Adcox:2001jp,Adcox:2004mh,Arsene:2004fa,Back:2004je,Adams:2005dq,Aamodt:2010jd,CMS:2012aa,Aad:2015wga,Adler:2002tq} that have been interpreted as evidence for strong partonic interactions within the dense medium that causes the quenching of jets. A direct observation of
this effect using jets was provided by ATLAS \cite{Aad:2010bu} and CMS
\cite{CMS_dijet2010, Chatrchyan:2012nia} through a comparison of the \pt balance of dijets in PbPb and pp collisions.
In head-on PbPb
collisions, a large increase in asymmetric dijet events was observed relative to the pp reference. This reflects the difference in energy lost by the two scattered partons in the medium, an effect that becomes more pronounced as the path lengths travelled by the partons and the energy density of the medium increase. In pPb collisions, no excess in unbalanced dijets was observed~\cite{Chatrchyan:2014hqa},
leading to the conclusion that the dijet imbalance does not originate from initial-state effects.
A wide range of models was proposed to accommodate the dependence of dijet data on the jet \pt and the centrality of the collision, \ie on the degree of  overlap of the two colliding nuclei~\cite{He:2011pd,Young:2011qx,Qin:2010mn,Casalderrey-Solana:2014bpa,Mehtar-Tani:2013pia,CasalderreySolana:2010eh}.
Further evidence for parton energy loss was found in studies of correlations between isolated photons and jets in PbPb events~\cite{Chatrchyan:2012gt},
where the unmodified isolated photon provides a measure of the initial parton momentum~\cite{Chatrchyan:2012vq}.

As energy is conserved in all interactions in the medium, parton energy loss does not imply the disappearance of energy, but its redistribution in phase space such that it is not recovered with standard jet finding clustering methods. The observed jet quenching naturally leads to questions of how the angular and \pt distributions of charged particles are modified by the energy loss of partons as they traverse the medium. A measurement of these spectra can provide information about the physical
processes underlying parton energy loss, which can yield insights into the properties of the
strongly interacting medium \cite{Kurkela:2014tla}. Particle distributions inside the jet cone (within
$\Delta = \sqrt{ \smash[b]{ (\eta_\text{trk}-\eta_\text{jet})^2 + (\phi_\text{trk}-\phi_\text{jet})^2 } } =  0.2$--0.4, where $\phi$ is the azimuthal
angle in radians and $\eta$ is the pseudorapidity) were studied in terms
of jet fragmentation functions and jet shapes~\cite{Chatrchyan:2012gw,Chatrchyan:2013kwa,Chatrchyan:2014ava,Aad:2014wha}.
These distributions show a moderate
softening and broadening of the in-cone fragmentation products in PbPb collisions compared
to pp data. However, the observed changes account for only a small fraction of the dijet momentum imbalance, indicating
that a large amount of energy is transported outside of the jet cone through interactions in the medium.

Identifying the distribution of particle \pt surrounding the jets (\ie the pattern of \pt ``flow'' relative to the dijet system) is challenging, as the ``lost'' \pt is only of
the order of 10\GeV, while the total \pt from soft processes forming the underlying event (UE) in a head-on (central) PbPb collision is about three orders of magnitude  larger~\cite{dNdeta,Chatrchyan:2012mb}. The angular distribution of the radiated energy is a priori unknown.
To overcome these difficulties, CMS previously used the ``missing \pt" method that exploits momentum conservation and azimuthal symmetry in dijet events. This method makes it possible to distinguish the correlated particles carrying the energy lost by jets from the uncorrelated particles, the directions of which are not related to the axes of the jets~\cite{CMS_dijet2010}. The momenta of all charged-particle tracks were therefore projected onto the jet direction, leading to a balancing of the uncorrelated particles, and thereby revealed the \pt flow relative to the dijet system. In pp events, imbalance in the \pt of leading and subleading jets is accommodated through three-jet and multijet final states. In PbPb collisions, quenching effects modify the spectrum and angular distribution of particles that recover the \pt balance within the dijet system. These studies
showed that the overall energy balance is restored only when low-momentum particles
($\pt \approx  0.5$--2\GeV) at large angles to the jet axis ($\Delta > 0.8$) are considered.

The original CMS analysis used a PbPb data sample corresponding to an integrated luminosity of 10\mubinv~\cite{CMS_dijet2010}, which was insufficient for a detailed study
of the angular pattern. In addition, no
pp data at the same collision energy was available at the time. In this paper, PbPb data corresponding to an integrated luminosity of 166\mubinv from a heavy-ion run at a nucleon-nucleon center-of-mass energy of 2.76\TeV, and pp data corresponding to an integrated luminosity of 5.3\pbinv taken at the same center-of-mass energy are used in a more comprehensive study.
The new data provide an opportunity for detailed characterization of the multiplicity, momentum, and angular distribution of particles associated with the flow of \pt in dijet events in PbPb and pp collisions, as a function of collision centrality and dijet \pt asymmetry. Collision centrality refers to configurations with different impact parameters of the lead nuclei. By changing centrality, the dependence of jet quenching can be studied as a function of the size and density of the medium.

To study the \pt flow relative to the dijet system, two complementary approaches are pursued, both relying on the cancellation of contributions from the uncorrelated UE.
First, the \pt of individual tracks are projected onto the dijet axis, defined as the bisector of the leading (highest \pt) jet axis and the subleading (next highest \pt) jet axis, with the latter flipped by $\pi$ in $\phi$.  These projections are then summed to investigate the overall $\pt$ flow in dijet events. This ``missing \pt'' analysis is used to
study how the lost momentum is distributed as a function of the
separation of the track from the jet axis, $\Delta$. The second approach involves the study of the difference in the total number of particles emitted in the leading  and subleading jet hemispheres. The measurements are carried out as a function of the collision centrality in PbPb collisions, and
as a function of the dijet \pt imbalance in pp and PbPb collisions. To investigate how differences in jet fragmentation affect energy loss mechanisms, jets are clustered using several anti-$\kt$ $R$ parameters (0.2, 0.3, 0.4 and 0.5) \cite{FastJet,bib_antikt}.

\section{CMS detector}

The central feature of the CMS apparatus is a superconducting solenoid with a 6\unit{m} internal diameter. Within the superconducting solenoid volume are a silicon pixel and strip tracker, a lead tungstate crystal electromagnetic calorimeter (ECAL), and a brass and scintillator hadron calorimeter (HCAL), each composed of a barrel and two endcap sections. Forward calorimeters extend the pseudorapidity~\cite{Chatrchyan:2008zzk} coverage provided by the barrel and endcap detectors. Muons are measured in gas-ionization detectors embedded in the steel flux-return yoke outside the solenoid.

The silicon tracker measures charged particles within the pseudorapidity range $\abs{\eta}< 2.5$. It consists of 1440 silicon pixel and 15\,148 silicon strip detector modules and is located in the 3.8\unit{T} field of the superconducting solenoid. For nonisolated particles of $1 < \pt < 10\GeV$ and $\abs{\eta} < 1.4$, the track resolutions are typically 1.5\% in \pt and 25--90 (45--150)\mum in the transverse (longitudinal) impact parameter \cite{TRK-11-001}. The ECAL has coverage up to $\abs{\eta}=1.48$, and the HCAL up
to $\abs{\eta}=3$.
Steel and quartz fibre hadron forward (HF) calorimeters extend the acceptance to \mbox{$\abs{\eta}=5$}.
For central $\eta$,
the calorimeter cells are
grouped in projective towers of granularity
$\Delta\eta\times\Delta\phi = 0.087 \times 0.087$.
The ECAL was initially calibrated using test beam electrons, and then
with photons from $\pi^0$ and $\eta$ meson decays and electrons
from \PZ boson decays~\cite{CMS-DP-2011-008,Khachatryan:2015iwa,Khachatryan:2015hwa}.
The energy scale in data agrees with that in the simulation to better than 1\,(3)\% in the barrel (endcap) region, $\abs{\eta}<1.5$ ($1.3<\abs{\eta}<3.0$) ~\cite{CMS-PAS-EGM-10-003}.
Hadron calorimeter cells in the $\abs{\eta}<3$ region are calibrated primarily with test beam data and
radioactive sources~\cite{hcal_jinst,Abdullin:2009zz}. A more detailed description of the CMS detector, together with a definition of the coordinate system and kinematic variables, can be found in Ref.~\cite{Chatrchyan:2008zzk}.

\section {Monte Carlo simulation}
\label{MCdefinition}

To study the performance of jet reconstruction in PbPb and pp collisions,
dijet events in nucleon-nucleon collisions are simulated with the \PYTHIA Monte Carlo (MC) event generator~\cite{Pythia} (version 6.423, tune Z2~\cite{Z2,Field}). To account for isospin effects present in PbPb collisions, the underlying pp, pn, and nn subcollisions are weighted by cross sections using the model from Ref.~\cite{Lokhtin:2005px}. For the simulation of dijet signals, a minimum hard-interaction scale of 30\GeV is used to increase the number of dijet events.

To model the PbPb UE, minimum bias PbPb events
are simulated with the \textsc{hydjet} event generator, version 1.8~\cite{Lokhtin:2005px}. The parameters of this version are tuned to reproduce total particle multiplicities, improve agreement with the observed charged-hadron spectra, and to approximate the fluctuations in UE seen in data. Proton-proton collisions are generated using leading-order (LO) {\PYTHIA} (without \textsc{hydjet} simulation). Full detector simulation using the
\GEANTfour{} package~\cite{bib_geant} and the standard CMS analysis chain are used to process both {\PYTHIA} dijet events and \PYTHIA dijet events embedded into \textsc{hydjet} events (denoted \textsc{pythia+hydjet} in this paper).

Jet reconstruction is studied using the jet information in the \PYTHIA generator in comparison to the same fully reconstructed jet in \textsc{pythia+hydjet}, after matching the generator-level and reconstructed jets in angular regions of $\Delta^{\rm reco,gen} = \sqrt{ \smash[b]{ (\eta^{\rm gen}_\text{jet}-\eta^{\rm reco}_\text{jet})^{2}+(\phi^{\rm gen}_\text{jet}-\phi^{\rm reco}_\text{jet})^{2} } } < R$.

\section{Jet reconstruction}
\label{sec:Jetreco}

Jet reconstruction in heavy-ion collisions at CMS is performed using the anti-$\kt$ algorithm and distance parameters $R = 0.2$ through $0.5$, encoded in the FastJet framework~\cite{FastJet}. Jets are reconstructed based on energies deposited in the CMS calorimeters. The probability of having a pileup collision is 23\%, and the average transverse energy ($E_{\rm T}$) associated with the UE is less than 1\GeV. For pp collisions, no subtraction is employed for the underlying event (UE) nor for pileup from overlapping pp interactions. Whereas, for PbPb collisions, a new ``HF/Voronoi" algorithm is used to subtract the heavy-ion background~\cite{CMS-DP-2013-018}. The transverse energy is defined by $E_{\rm T} = \sum E_{i} \sin{(\theta_{i})}$, where $E_{i}$ is the energy of the ${i}^{\rm th}$ particle in the calorimeter, $\theta_{i}$ is the polar angle of particle $i$ measured from the beam axis, and the sum is over all particles emitted into a fixed $\Delta$ in an event.

The HF/Voronoi algorithm removes the UE contribution by estimating the $E_{\rm T}$ contribution from the UE at central $\eta$, and its azimuthal dependence, from deposition in the HF detector. The estimation is performed using a polynomial model that is trained using a singular-value decomposition method~\cite{GolubReinschSVD}, separately on minimum bias data and MC simulation. After an average $E_{\rm T}$ is subtracted from each calorimeter tower, based on its location in $\eta$ and $\phi$, the calorimeter towers containing non-physical negative $E_{\rm T}$ are evened out by redistributing the energy in neighboring positive $E_{\rm T}$ towers in a circular region of the parameter $R + 0.1$. The redistribution is implemented by minimizing a metric that describes the total energy difference before and after the process, given that after the redistribution all towers have positive energy.

The initial calorimetric $E_{\rm T}$ values are corrected as a function of \pt and $\eta$ to match the jets clustered using all particles, except neutrinos, at the generator level of \PYTHIA. The consistency of the corrected jet energy scale (JES), defined as $\langle \pt^{\rm reco}/\pt^{\rm gen} \rangle$, is checked as a function of $\pt$ and $\eta$ using \textsc{pythia+hydjet} events in bins of event centrality. The deviations are within 2\% for all centrality, \pt, and $\eta$ bins, and less than 1\% for jet \pt greater than 60\GeV.

The nonlinear response of the calorimeter as a function of particle energy gives jets that fragment into many particles with smaller energies a smaller  response relative to the jets of same energy but with fewer fragments. To account for the dependence of JES on the fragmentation of jets, an additional correction is applied as a function of reconstructed jet \pt, and as a function of the number of charged particles with $\pt > 2\GeV$ in a cone of $R$ around the jet axis. The number of charged particles in \textsc{pythia+hydjet} is calculated using the \pt values obtained after the ``HF/Voronoi'' subtraction. For \PYTHIA, \pt values without any UE subtraction are used to calculate the number of charged particles. The fragmentation-dependent correction applied in PbPb collisions is calculated using \textsc{pythia+hydjet} events with matching UE activity. This correction results in a reduction in separation of the JES for quark and gluon jets, and also lessens the impact of jet reconstruction on fragmentation of the leading and subleading jets.

The residual JES that accounts for the difference in calorimeter response in data and MC events is calculated using dijet balance in pp and peripheral (50--100\% centrality) PbPb collisions~\cite{Chatrchyan:2011ds}, based on data. This difference is found to be less than 2\% for $\abs{\eta} < 2$.

\section{Track reconstruction}
\label{sec:Trackreco}

For studies of pp data and \PYTHIA MC events, charged particles are reconstructed using the same iterative method~\cite{TRK-11-001} as in previous CMS analyses of pp collisions. However, for PbPb data and \textsc{pythia+hydjet} events, a different iterative reconstruction ~\cite{CMS:2012aa,Chatrchyan:2013kwa} is employed after extending the global tracking information down to $\pt = 0.4\GeV$. To minimize the impact of inefficiencies in track reconstruction caused by the \pt resolution in track seeds near the 0.4\GeV threshold, only tracks with $\pt > 0.5\GeV$ are used in this analysis.

Reconstructed tracks in \PYTHIA and \textsc{pythia+hydjet} simulations are matched to primary particles using the associated hits, \ie, charged particles that are produced in the interaction or are remnants of particles with a mean proper lifetime of less than $5 \times 10^{13}\GeV^{-1}$. The misidentification rate is defined as the fraction of reconstructed tracks that do not match any charged particle (primary or otherwise). The multiple reconstruction rate is given by the fraction of primary particles that are matched to more than one reconstructed track. Tight track quality criteria are applied to reduce the rate of misidentified or secondary particles~\cite{TRK-11-001}. Requirements are less restrictive for pp than for PbPb collisions. Heavy-ion tracking requires a larger number of hits in the tracker and a smaller normalized fit $\chi^{2}$ value for fits to reconstructed tracks. For both systems, tracks are required to be compatible with the vertex with the largest value in the sum of their \pt.

In pp collisions, the track reconstruction efficiency is  ${\approx} 90\%$ at $\pt = 10\GeV$ and 80\% at 0.5\GeV. The misidentification rate for tracks is ${<}2\%$ for $\pt > 1\GeV$ and slightly higher below this value.
The contribution from secondary particles is subtracted, as the secondary-particle rate is as high as 2\%. The multiple reconstruction rate is smaller than 1\%. The efficiency and misidentification corrections are calculated as a function of $\eta$, $\phi$, $\pt$, and the distance to the nearest jet axis, while simpler secondary-particle and multiple reconstruction corrections are applied that depend only on the $\eta$ and $\pt$ values of charged particles.

As the track reconstruction efficiency in pp collisions is larger than in PbPb collisions, the momentum flow can be measured with higher precision, while in the high-multiplicity environment of heavy-ion collisions track reconstruction remains a challenge. In PbPb collisions, the reconstruction efficiency for primary charged particles, after implementing the above track quality criteria, is approximately 70\% at $\pt \approx 10\GeV$. Efficiency  starts to drop for $\pt$ below 5\GeV and at 0.5\GeV the efficiency is 30\%. The misidentification rate for tracks with $\pt = 0.5\GeV$ is ${\approx} 35\%$, but decreases to values smaller than $2\%$ and for $\pt > 1\GeV$. The secondary-particle rate and multiple-reconstruction rate are, respectively, less than 0.5\% and 0.3$\%$ over the whole \pt range in the analyis. No corrections are applied for these in PbPb collisions. Using \textsc{pythia+hydjet} simulations, track reconstruction efficiency and misidentification rates are evaluated as a function of the $\eta$, $\phi$, and \pt of the track, as well as the centrality of the collision, and the smallest distance in $\Delta$ between the track and a jet with $\pt > 50\GeV$.

Tracks used in the analysis are weighted with a factor to correct for the effects described above. The value of this correction is
\begin{equation}
{c}^\text{trk} = \frac{(1-\text{misreconstruction}) \, (1-\text{secondary-particle})}{(\text{efficiency})\,(1+\text{multiple-reconstruction})},
\label{Eqn:ctrk}
\end{equation}
where secondary-particle and multiple-reconstruction rates are set to zero for PbPb collisions.

\section{Analysis}
\label{sec:analysis}

Events are selected using an inclusive single-jet trigger with jet \mbox{$\pt > 80\GeV$}. To suppress electronic noise, cosmic rays, and beam backgrounds, events are required to satisfy selection criteria documented in refs~\cite{CMS_dijet2010,Chatrchyan:2012gt}. Events passing selections are subject to offline jet reconstruction. To select samples containing high-\pt dijets, events are required to have a leading (subleading) jet in the range of $\abs{\eta}<2$ with a
corrected jet $\pt> 120\, (50)\GeV$. The single-jet trigger is fully efficient for events with the requirement on the leading jet \pt for all the $R$ parameters in the analysis. To select a dijet topology, the azimuth between the leading and subleading jets is required to be $\Delta\phi_{1,2}  = \abs{\phi_{1} - \phi_{2}} > 5\pi/6$. Once leading and subleading jets are identified within the initial range of $\abs{\eta}<2$, both jets are then restricted to be within a tighter $\abs{\eta}$. For measurements that offer comparison to a previous analysis~\cite{CMS_dijet2010}, we use the previous selection of $\abs{\eta}<1.6$. For those that extend up to large angular distances $\Delta$, a tighter requirement of $\abs{\eta}<0.6$ is applied, such that leading and subleading jets are far from the edge of the tracker and all ranges in $\Delta$ fall within the acceptance.

This analysis aims to provide information that would aid the characterization of the energy loss mechanisms responsible for the increase in the fraction of unbalanced dijet pairs in central PbPb relative to pp collisions. As hard-scattered partons travel and shower in the QGP, they can both trigger a coherent medium response and undergo interactions in the medium that modify the showers of both partons. However, the enhancement in unbalanced dijet pairs suggests that, on average, the subleading jet loses more energy than the leading jet. The modification in jet balance must be compensated by the remaining, unclustered constituents of the event, as each interaction conserves overall momentum.

The particles that provide the \pt balance are correlated with the jet axes, but the particles that are not affected by the interaction of the partons with the medium are evenly distributed in azimuth relative to the individual directions of the leading and subleading jets. The total \pt of these particles is uncorrelated with the dijet pair. To differentiate the uncorrelated and correlated particles, we compare differences in multiplicity in leading and subleading jet hemispheres. In addition, we measure modifications in the \pt spectrum of charged particles that contribute to the overall \pt balance in the event, as well as their angular distribution with respect to the dijet system. Using the azimuthal symmetry of the jet axes relative to the UE makes it possible to perform precise measurements for particles down to $\pt = 0.5\GeV$, and angles as large as $\Delta = 1.8$. This provides constraints on energy loss mechanisms despite the small signal-to-background ratio.

The cancellation of the uncorrelated UE depends on azimuthal symmetry of the areas selected around the leading and subleading jets relative to the axis of projection. As mentioned above, to ensure this requirement, the dijet azimuthal angle ($\phi_\text{dijet}$) is defined as the average $\phi$ of the leading and subleading jets after the subleading jet is reflected around the origin. In contrast with previous publications~\cite{CMS_dijet2010}, $\phi_\text{dijet}$ is preferred over $\phi_{1}$ (the $\phi$ of the leading jet) for the projection axis, because the latter choice breaks azimuthal symmetry, by generating particles near the leading jet that have larger projections at small angles relative to particles produced at the same distance to the subleading jet.

The perfect cancellation of contributions from particles
to \pt flow, and to differences in hemisphere multiplicities from UE, take place only when there is no interaction between UE and the jets. This is the case in \textsc{pythia+hydjet} simulations. In data, due to the variations in path length in medium traversed by jets there are complicated correlations between particles from different interactions and jet directions. These correlations comprise a part of the signal probed in this analysis.

The observables used in this analysis are measured in bins of centrality and dijet imbalance. The dependence on centrality in PbPb collisions is investigated in terms of the emergence and enhancement of jet quenching effects as the size of the medium and energy density increase, and the dijet imbalance enriches events with subleading jets that lose more energy than the leading jet.  To define centrality classes, collisions with inelastic hadronic interactions are divided into percentages according to the  $E_{\rm T}$ of calorimeter towers summed in the HF, and events are assigned into classes of centrality based on these total sums in the HF. The distribution in this $E_{\rm T}$ is used to divide the event sample into bins, each representing 0.5\% of the total nucleus-nucleus interaction cross section. Following Refs.~\cite{CMS_dijet2010,Chatrchyan:2012nia}, we quantify \pt imbalance through the asymmetry ratio $ \AJ = (p_{{\rm T},1}-p_{{\rm T},2})/(p_{{\rm T},1}+p_{{\rm T},2})$, where $p_{\rm{T},1}$ and $p_{\rm{T},2}$ are the \pt of the leading and subleading jets within $\eta < 2.0$, respectively.  The $\AJ$ boundaries used in the analysis are 0.11, 0.22, 0.33 and 0.44, which correspond to $p_{\rm T,2}/p_{\rm T,1}$ values of 0.8, 0.64, 0.50 and 0.42, respectively.

\subsection{Difference in multiplicities}

The events are bisected with a plane perpendicular to $\phi_\text{dijet}$ into two hemispheres associated with the leading and subleading jets. The multiplicity difference is defined as the difference between the corrected number of tracks with $\pt > 0.5\GeV$ ($N_\text{trk}^\text{Corrected} = \sum {c}^\text{trk}$) in these two hemispheres:
\begin{equation}
\Delta_\text{mult} = N_\text{trk}^\text{Corrected}|_{|\phi_\text{trk} - \phi_\text{dijet}|>\pi/2} - N_\text{trk}^\text{ Corrected}|_{\abs{\phi_\text{trk} - \phi_\text{dijet}}<\pi/2}.
\end{equation}

Positive $\Delta_\text{mult}$ means that an excess of particles is found in the hemisphere of the subleading jet, relative to the number of particles in the leading jet hemisphere. This quantity is measured event-by-event and then averaged in bins of the observables of interest. It is sensitive to the number of jets in a given hemisphere and their fragmentation, as well as to the additional particles produced in jet quenching or through some specific response of the QGP medium in one of the two hemispheres.

To select events that show consequences of jet quenching, the measurement is carried out as a function of $\AJ$ and collision centrality. The $\AJ$-dependent measurement is performed for jets with a distance parameter of $R = 0.3$.

To see modifications in the \pt spectrum associated with the difference in multiplicities in two hemispheres, $\Delta_\text{mult}$ is measured for track \pt ranges of 0.5--1, 1--2, 2--4, 4--8, and 8--300\GeV, and divided by the bin width. The measurement is repeated for different $R$ parameters.

To be consistent with the measurement of \pt balance, the leading and subleading jets used in the $\AJ$-dependent $\Delta_\text{mult}$ measurement are required to fall in the pseudorapidity region of $\abs{\eta}<1.6$. The leading and subleading jets used in the $R$-dependent measurement are required to be within $\abs{\eta}<0.6$. Although in both cases jets with $\abs{\eta} > 2$ are excluded, it is important to note that starting jet reconstruction with a cutoff $\abs{\eta}<1.6$, (or $<0.6$) is different than using the $\abs{\eta}<2$ selection for determining the highest-\pt jets and then applying a tighter requirement, since events in which the leading or subleading jets are found in the range between $\abs{\eta} = 1.6$ (or $0.6$) and $\abs{\eta} = 2.0$ are also excluded.

\subsection{Transverse momentum balance}
\label{sec:momentumBalance}

Detailed information about the \pt flow relative to the dijet system can be obtained by studying the contribution of tracks to the overall \pt balance in the event, as characterized by individual track \pt and angle relative to the jets. To calculate the \pt balance, the \pt of tracks are projected onto the dijet axis. For each track, this projection is defined as
\begin{equation}
 \pt^{\parallel} =
 -{c}^\text{trk} \, \pt^\text{trk} \, \cos{(\phi_\text{trk} - \phi_\text{dijet})},
\end{equation}
where, as mentioned in Section~\ref{sec:Trackreco}, the correction for reconstruction effects  accounts for the misreconstruction rate and reconstruction efficiency for PbPb collisions, with values specified by Eq.~\ref{Eqn:ctrk}.
In addition, secondary particle and multiple reconstruction rates are corrected in pp collisions.

Particles that make a positive contribution in $\Delta_\text{mult}$ also have positive $\pt^{\parallel}$, as the cosine function changes sign at $\pi/2$. These two observables therefore map onto each other with a weight in track $\pt$ and $\cos{(\phi_\text{trk} - \phi_\text{dijet})}$.

To study the angular recovery rate (rate at which imbalance is restored, as momentum contributions are included further from the jet cone) and the associated spectra of \pt balance, tracks that fall in annular regions around the jet axes are grouped together according to their \pt. In each event, $\pt^{\parallel}$ values of these group of tracks are summed to obtain $\PTslash^{\parallel}$. For each region, $\PTslash^{\parallel}$ is calculated in track \pt ranges of 0.5--1, 1--2, 2--4, 4--8, and 8--300\GeV.
Annular regions are defined in $\Delta = \sqrt{ \smash[b]{ (\phi_\text{trk} - \phi_\text{jet})^2 + (\eta_\text{trk} - \eta_\text{jet})^2} } $ and binned between $\Delta=0.0$--1.8 in steps of 0.2. In addition, the contribution from charged particles that fall outside of this range are all collected in an extra overflow bin. These particles lie in the range of $1.8< \Delta < 3.6$, depending on the $\eta$ of the dijet pair. No anti-$\kt$ clustering is employed in the calculation of $\Delta$, and tracks are defined to lie within circular regions in pseudorapidity and azimuth. The axes used to define the annuli differ from the projection axis, $\phi_\text{dijet}$. For large $\Delta$, the
annuli around the leading and subleading jets can overlap, in which case, the track used in the overlap region when calculating $ \PTslash^{\parallel}$, is the one in the annulus at smaller radius. The overlaps do not occur before $\Delta = 5\pi/12$.

The $\PTslash^{\parallel}$ is averaged over events with a specific $\AJ$ value separately for pp and PbPb collisions, and for PbPb collisions they are divided into classes of collision centrality. This average is denoted as $\langle \PTslash^{\parallel}\rangle_{\pt^\text{trk},\Delta}$, to indicate that within each event the balance is calculated using a subset of tracks with specific $\Delta$ and \pt.

Using the track $\pt$ and $\Delta$ parameters limits the selections on collision centrality and $\AJ$  because of the statistical imprecision of the data. For more detailed analysis of the dependance of track $\pt$ on event properties, $\Delta$ binning can be removed by adding up the \MPTpTdR values for each $\Delta$ bin, which is identical to not having annular requirements in the first place, to obtain
\begin{equation}
\langle \PTslash^{\parallel} \rangle_{\pt^{\text{trk}}} =
 \sum\limits_{\Delta} \langle \PTslash^{\parallel} \rangle_{\pt^{\text{trk}},\Delta}.
 \label{Eqn:four}
\end{equation}

The \pt balance, as in Eq.~\ref{Eqn:four}, calculated for tracks in a given \pt range usually yields nonzero values, because of the differences in \pt spectra of particles in subleading jet hemisphere relative to the spectra in the leading jet hemisphere. Summing the signed \MPTpT values for each track \pt bin provides an overall \pt balance in the event for tracks with $0.5 < \pt < 300\GeV $, that takes values close to zero, because of momentum conservation. There can still be a deviation from zero because of the particles with  $\pt < 0.5\GeV$, as well as for those particles that fall out of the tracker coverage in pseudorapidity that are not included in the measurement. This sum corresponds to
\begin{equation}
\langle \PTslash^{\parallel} \rangle_{\Sigma} =
 \sum\limits_{p_{\rm{T}}^{\text{trk}}} \langle \PTslash^{\parallel} \rangle_{\pt^{\text{trk}}}.
\end{equation}

The angular distribution of \pt balance is studied differentially in bins of track \pt by $\langle \PTslash^{\parallel}\rangle_{\pt^\text{trk},\Delta}$, as described above, and adding up the contribution from different track \pt bins gives
\begin{equation}
\langle \PTslash^{\parallel} \rangle_{\Delta} =
 \sum\limits_{p_{\rm{T}}^{\text{trk}}} \langle \PTslash^{\parallel} \rangle_{\pt^{\text{trk}},\Delta},
\end{equation}

which defines the contribution of all tracks with $0.5 < \pt < 300\GeV$ in a given annulus to total \pt balance. This \MPTdR, summed over all $\Delta$ intervals, yields $\langle \PTslash^{\parallel} \rangle_{\Sigma}$. Instead of summing all $\Delta$ bins, to calculate the recovery of balance as radius gets larger, the annuli can be summed from $\Delta = 0$ up to the angle of interest, and a cumulative balance inside a cone calculated, as
\begin{equation}
\langle \PTslash^{\parallel} \rangle_{[0,\Delta]} =
 \sum\limits_{\Delta^\prime = 0}^{\Delta^\prime = \Delta} \langle \PTslash^{\parallel} \rangle_{\Delta^\prime}.
\end{equation}

As mentioned previously, for consistency with the analysis in Ref.~\cite{CMS_dijet2010}, in calculations that integrate over $\Delta$ , \eg for, \MPTpT and \MPTsum, only events in which both leading and subleading jets fall within
$\abs{\eta}<1.6$ are included in the measurement of \pt balance. For measurements where contributions of different annuli are studied, to ensure full tracker coverage around jets over $\Delta < 1.8$ for  \MPTpTdR, \MPTdR, and \MPTcum,  tighter restrictions are required on the pseudorapidity of leading and subleading jets ($\abs{\eta}<0.6$) after they are found within  $\abs{\eta}<2$.

\section{Systematic uncertainties}
\label{sec:systematics}

\begin{table*}[!ht]
\centering
\topcaption{Systematic uncertainties in \MPTdR for jets clustered with distance parameter of 0.3 in pp, and in central and peripheral PbPb collisions, for different $\AJ$ selections. Uncertainties are shown as shifts in the values in units of\GeV (rather than as fractions) for two $\Delta$ selections.}
\label{table:SysAdepMpt}

\cmsTableResize{
\begin{tabular}{l|xc|xc|xc}

\multicolumn{1}{c}{} & \multicolumn{6}{c}{Values integrated over $\AJ$} \\
\cline{2-7}
\multicolumn{1}{c}{}  & \multicolumn{2}{c}{pp} &  \multicolumn{2}{|c|}{PbPb, 30--100\%} & \multicolumn{2}{c}{PbPb, 0--30\%} \\
\hline
$\Delta$ &{<}0,2 & 0.2--2.0 &{<}0,2 & 0.2--2.0 &{<}0,2 & 0.2--2.0 \\
\hline
Jet reconstruction &{<}1 & 0.0--0.2 & 1 & 0.1--0.2 & 1 & 0.1--0.4 \\
Data/MC differences for JES & 1 & 0.1--0.2 & 2 & 0.1--0.3   & 2 & 0.1--0.3 \\
Fragmentation dependent JES &{<}1 & 0.1--0.2 & 2 & 0.1--0.2 & 1 & 0.1--0.4  \\
Track corrections &{<}1 & ${<}0.1$ & 1 & 0.0--0.2  & 2 & 0.2--0.9 \\
Data/MC differences for tracking& 1 & 0.0--0.1  & 1 & 0.1--0.2 & 1 & 0.1--0.2 \\
\hline
Total & 1 & 0.1--0.3 & 2 & 0.2--0.3 & 3 & 0.2--1.0 \\
\hline

\multicolumn{7}{c}{} \\
\multicolumn{1}{c}{} & \multicolumn{6}{c}{$\AJ < 0.22$} \\
\cline{2-7}
\multicolumn{1}{c}{}  & \multicolumn{2}{c}{pp} &  \multicolumn{2}{|c|}{PbPb, 30--100\%} & \multicolumn{2}{c}{PbPb, 0--30\%} \\
\hline
$\Delta$ &{<}0,2 & 0.2--2.0 &{<}0,2 & 0.2--2.0 &{<}0,2 & 0.2--2.0 \\
\hline
Jet reconstruction & {<}1 & 0.1--0.2  & 1 & 0.1--0.2  & 1 & 0.1--0.4 \\
Data/MC differences for JES & 1	& 0.1--0.2 & 2 & 0.1--0.4 & 2 & 0.2--0.4 \\
Fragmentation dependent JES & {<}1	& 0.1 & 2 & 0.1--0.4  & 1 & 0.1--0.5  \\
Track corrections& {<}1 	& $<$0.1 & 1 & 0.1 &	2 & 0.1--0.6 \\
Data/MC differences for tracking& {<}1 & 0.0--0.1 & 1 & 0.1  & 1& 0.1 \\
\hline
Total & 1 & 0.1--0.3 & 2 & 0.2--0.4 & 3 & 0.2--0.6 \\
\hline

\multicolumn{7}{c}{} \\
\multicolumn{1}{c}{} & \multicolumn{6}{c}{$\AJ > 0.22$} \\
\cline{2-7}
\multicolumn{1}{c}{}  & \multicolumn{2}{c}{pp} &  \multicolumn{2}{|c|}{PbPb, 30--100\%} & \multicolumn{2}{c}{PbPb, 0--30\%} \\
\hline
$\Delta$ &{<}0,2 & 0.2--2.0 &{<}0,2 & 0.2--2.0 &{<}0,2 & 0.2--2.0 \\
\hline
Jet reconstruction &2 & 0.1--0.5  & 1 & 0.1--0.6  & 2	& 0.2--0.6 \\
Data/MC differences for JES  & 2	& 0.1--0.3& 3 & 0.2--0.5 & 3 & 0.3--0.6\\
Fragmentation dependent JES& 1	& 0.1--0.5 & 1 & 0.1--0.7  & 1 & 0.2--0.6  \\
Track corrections & {<}1 & 0.1 & 1 & 0.1--0.3  & 3 & 0.2--1.1\\
Data/MC differences for tracking& 2 & 0.1--0.2 & 2	& 0.1--0.2 & 2 & 0.1--0.3   \\
\hline
Total& 3	& 0.3--0.8 & 3 & 0.3--0.9  & 4 & 0.4--1.4 \\
\hline
\end{tabular}
}
\end{table*}

The sources of major systematic uncertainty can be categorized into two groups; biases related to jet reconstruction and those related to track reconstruction. Effects associated with event selection and beam background rejection are found to be negligible.

The biases related to jet reconstruction are caused by smearing of jet \pt due to energy resolution and uncertainties in the JES. These factors can change the \pt-ordering of jets in the event, resulting in the interchanging of leading and subleading jets, or causing third jet to replace the subleading jet. The uncertainties are estimated as a function of centrality and $\AJ$ in each charged-particle \pt range, using \PYTHIA and \textsc{pythia+hydjet} simulations to compare observables calculated with reconstructed jets to generator-level jets.  A bin-by-bin correction is applied to data to account for the observed jet reconstruction bias. This uncertainty includes the effect of jet-angular resolution. However, the size of the bins in the $\Delta$-dependent measurement is significantly larger than a typical angular resolution, which therefore has a negligible effect on the observables. Going from $R=0.2$ to $0.5$, the angular resolution, defined by the standard deviation of the $\Delta^\text{reco,gen}$ distribution, increases from 0.020 to 0.025 for leading jets, and from 0.025 to 0.035 for subleading jets in pp. The same holds in 30--100\% centrality PbPb collisions. In the most central 0--30\% of events, the corresponding ranges are 0.020--0.035 and 0.025--0.045, respectively.

After implementing the fragmentation-dependent jet energy corrections there is up to 5\% difference between the JES for quark and gluon jets at $\pt < 50\GeV$, and the difference disappears for high-\pt jets. Additional checks are therefore pursued to account for possible discrepancies in the performance of  jet energy corrections in data and in MC simulations. A modification in flavor content of jets due to quenching can lead to an under- or over-correction of the jet energy in data. Also, the uncertainty in the JES from differences in simulation and detector conditions is calculated to be 2\% using a data-based ``tag-and-probe" technique that depends on dijet balance in a control sample of peripheral PbPb events~\cite{Chatrchyan:2011ds}. The jet \pt is changed up and down for leading and subleading jets in an asymmetric manner (leading JES is increased while subleading JES is decreased) as a function of jet \pt, to account for the differences in JES between quark and gluon jets and the data-based JES uncertainty. Because the number of charged particles is a parameter in these corrections, and can make the fragmentation-dependent jet energy corrections sensitive to quenching effects, the difference in the observables before and after corrections in MC events is compared to the corresponding change in data, and the discrepancy between data and simulation is quoted as an additional source of uncertainty.

Uncertainties related to track reconstruction are calculated in \PYTHIA and \textsc{pythia+hydjet} by comparing the results with generator-level charged particles to those with reconstructed tracks, after applying the track corrections discussed in Section~\ref{sec:Trackreco}. The small uninstrumented regions in the detector, and the correlation between track reconstruction efficiency and JES are the main causes of discrepancies observed between results with generator-level particles and reconstructed tracks.
The track corrections account for the inefficiencies due to uninstrumented regions. However, the bins used in $\eta$ and $\phi$ to calculate the reconstruction efficiency are larger than the size of the uninstrumented regions, and as a result cannot completely correct the effect of these.
An additional uncertainty is therefore added to account for the effect of differences in detector conditions and simulation of track reconstruction. This is achieved using the ratio of corrected to initial track \pt and $\eta$ spectra in data and simulations that are compared as track-quality selections are changed. The difference is found to be less than 5\%, which is included in the systematic uncertainty.

To calculate the total uncertainty, the uncertainties from sources mentioned above are summed in quadrature. The contribution of each item is summarized in Tables~\ref{table:SysAdepMpt}--\ref{table:SysRdepMpt} for the \MPTdR measurement. The systematic sources are given in terms of shifts in the value of each observable in a given bin in units of\GeV instead of \% changes, as the \MPTdR can vanish and can take values arbitrarily close to zero. Typically, \MPTdR is between 15--40\GeV near the jet axes ($\Delta < 0.2$), and less than 10\GeV at larger angles.

The dependence of uncertainties in dijet asymmetry and centrality is summarized in Table~\ref{table:SysAdepMpt} for jets with a distance parameter $R=0.3$. The jet energy resolution, can cause events to move across the $\AJ$ boundaries. Moreover, it is more likely for the leading jet in a highly imbalanced dijet event to be located in a region of an upward UE fluctuation in PbPb collisions. For these reasons, uncertainties related to jet reconstruction are larger in imbalanced dijet events. For well-balanced events, the uncertainty is comparable to that in the inclusive $\AJ$ selection, because the increase in effects from jet energy resolution balances the reduction of effects related to UE fluctuations.
Uncertainties in track reconstruction are larger in imbalanced than in balanced events, because of the correlation of track reconstruction efficiency and reconstructed jet energy. When a high-\pt track that carries a significant fraction of jet \pt is not reconstructed, the jet energy is under-corrected, and vice versa, the energy is over-corrected in events where the high-\pt track is found, because jet energy corrections are obtained for the average case where the high-\pt track might not be reconstructed. Events with highly imbalanced dijets can result from miscalculated jet energies caused by inefficiencies in track reconstruction.
Centrality of PbPb collisions does not affect the uncertainties within the jet cone as much as at larger angles, where the signal-to-background ratio gets smaller. Track and jet reconstruction uncertainties, caused by over-correction of the leading jet \pt because of upward UE fluctuations, in particular, tend to increase in central collisions. Uncertainties are smaller in pp than in PbPb collisions because of the absence of a heavy-ion UE, and differences in jet and track reconstruction that provide better measurement of jet \pt, larger track reconstruction efficiency, and lower track misidentification rates.

Uncertainties for small $\Delta$ are dominated by charged particles with $\pt > 8\GeV$, while at larger $\Delta$, low-\pt particles make up a larger fraction of the total uncertainty in events when there is no selection made on charged-particle \pt. The contribution from each range of track \pt to the uncertainty in \MPTdR, in other words the uncertainty in \MPTpTdR, is shown in Table~\ref{table:SysPtDep} for $R = 0.3$, in events with 0--30\% central PbPb collisions. Finally, as shown in Table~\ref{table:SysRdepMpt}, uncertainties in jet reconstruction and track reconstruction in MC events increase together with increasing $R$, as the UE inside the jet cone gets larger. However, JES difference between quark and gluon jets is smaller for large $R$ parameters, and uncertainties that account for JES differences in data and in MC events therefore decrease.

\begin{table*}[!ht]
\centering
\topcaption{Systematic uncertainties in \MPTpTdR in 0--30\% PbPb collisions, for jets clustered with a distance parameter of 0.3, as a function of charged-particle \pt. Uncertainties are shown as shifts in the values in units of\GeV (rather than as fractions) for two $\Delta$ selections.}
  \label{table:SysPtDep}

 \cmsTableResize{
 \begin{tabular}{l|xc|xc|xc}
\multicolumn{1}{c}{}  & \multicolumn{2}{c|}{$0.5 < \pt < 2\GeV$ } & \multicolumn{2}{c|}{$2 < \pt < 8\GeV$ } & \multicolumn{2}{c}{$ \pt > 8\GeV$ } \\
\hline
$\Delta$ &{<}0,2 & 0.2--2.0 &{<}0,2 & 0.2--2.0 &{<}0,2 & 0.2--2.0 \\
\hline
Jet reconstruction &  0,04	& 0.06--0.25 & 0,13 &	0.04--0.14 & 0,85 & 0.01--0.07\\
Data/MC differences for JES &	0,14 & 0.07--0.24 & 0,42 & 0.03--0.11 & 0,97	& 0.01--0.12 \\
Fragmentation dependent JES &  0,03 & 0.10--0.14 & 1,1 & 0.05--0.23 & 0,19 & 0.02--0.06\\
Track corrections & 0,09 & 0.08--0.64 & 0,27 & 0.06--0.13 &	1,78 & 0.01--0.07\\
Data/MC differences for tracking & 0,04 &	0.03--0.08 & 1,2 & 0.01--0.05 &	1,16 & 0.00--0.02\\
\hline
Total & 0,17 & 0.20--0.69 &	1,1 & 0.11--0.29 & 2,3 & 0.04--0.10\\
\hline
\end{tabular}
}
\end{table*}

\begin{table*}[!ht]
\centering
\topcaption{Systematic uncertainties in \MPTpTdR in 0--30\% PbPb collisions are shown for jets clustered with distance parameters of 0.2, 0.4 and 0.5. Uncertainties are shown as shifts in the values in units of\GeV (rather than as fractions) for two $\Delta$ selections.}
  \label{table:SysRdepMpt}

 \cmsTableResize{
 \begin{tabular}{l|xc|xc|xc}
 \multicolumn{1}{c}{} & \multicolumn{2}{c|}{$R = 0.2$} & \multicolumn{2}{c|}{$R = 0.4$} & \multicolumn{2}{c}{$R = 0.5$} \\
\hline
$\Delta$ &{<}0,2 & 0.2--2.0 &{<}0,2 & 0.2--2.0 &{<}0,2 & 0.2--2.0 \\
\hline
Jet reconstruction & 1 &  0.1--0.4  & 1 & 0.1--0.5 & 1 & 0.1--0.7 \\
Data/MC differences for JES & 2 & 0.1--0.5 & 2 & 0.1--0.4 & 2 & 0.1--0.3 \\
Fragmentation dependent JES & 1 & 0.1--0.4  & 1 & 0.1--0.3 & 1 & 0.1--0.3 \\
Track corrections & 2 & 0.2--0.7 & 2 & 0.1--1.1 & 2 & 0.1--1.1 \\
Data/MC differences for tracking & 1 & 0.1--0.2 & 1 & 0.1 & 1 & 0.1 \\
\hline
Total & 3 & 0.2--0.9 & 3 & 0.3--1.1  & 3 & 0.2--1.1\\
\hline
\end{tabular}
}
\end{table*}

 Although uncertainties in differences in multiplicities are calculated separately, their values are not listed in a table, because they can be approximated from the uncertainties in $\langle\PTslash^{\parallel}\rangle$ divided by the average charged particle \pt in that range. In 0--10\% central events, for $R = 0.3$, the dominant source is jet reconstruction, with an uncertainty caused by an upward fluctuation in the background under the leading jet, which is followed by the uncertainty in track reconstruction, and residual track reconstruction in data and in MC events that change by 0.5--1.5 particles, as a function of $\AJ$. The uncertainties increase with $R$ and with centrality from peripheral to central collisions.

\section{Results}
\label{sec:results}

\subsection{Dependence of the \texorpdfstring{\pt}{pT} balance in pp and PbPb on opening angles around jets}
\label{sec:results_AngDep}

Angular distribution of the \pt relative to the axis defined by the parton direction is a key for studying QCD processes responsible for parton energy loss. In models, large-angle modifications in the event due to jet quenching have been accommodated qualitatively through a response triggered in the hydrodynamic medium by the deposited energy~\cite{Tachibana:2014lja} and through the cascade of gluons created in medium-induced radiation processes~\cite{Blaizot:2014rla,Iancu:2015uja,Fister:2014zxa,Blaizot:2014ula}. Moreover, some MC implementations of jet quenching that modify partonic showers in \PYTHIA, such as \textsc{Q-pythia}, can generate soft particles at angles $\Delta > 0.8$, but this treatment modifies the fragmentation functions more severely than found in data~\cite{Apolinario:2012si,Armesto:2009fj}. Angular scales for different jet quenching mechanisms in perturbative QCD are related to momentum scales through time evolution of partonic interactions~\cite{Kurkela:2014tla}. Especially for QCD cascades in a sufficiently large medium, angular broadening is independent of the path length, and this mechanism might therefore produce a cumulative effect even after taking averages over different events where jets travel different path lengths in the QGP. The medium response may not have the same correlation between angular and momentum scales. The relative importance of each mechanism is unknown. Measuring the \pt spectra of $\langle \PTslash^{\parallel} \rangle$ as a function of $\Delta$ from the jet axis, denoted as \MPTpTdR, as discussed in Section~\ref{sec:analysis}, can provide information on the momentum scales at which certain quenching mechanisms become dominant.

\begin{figure}[h!t]
\centering
       \includegraphics[width=0.95\textwidth]{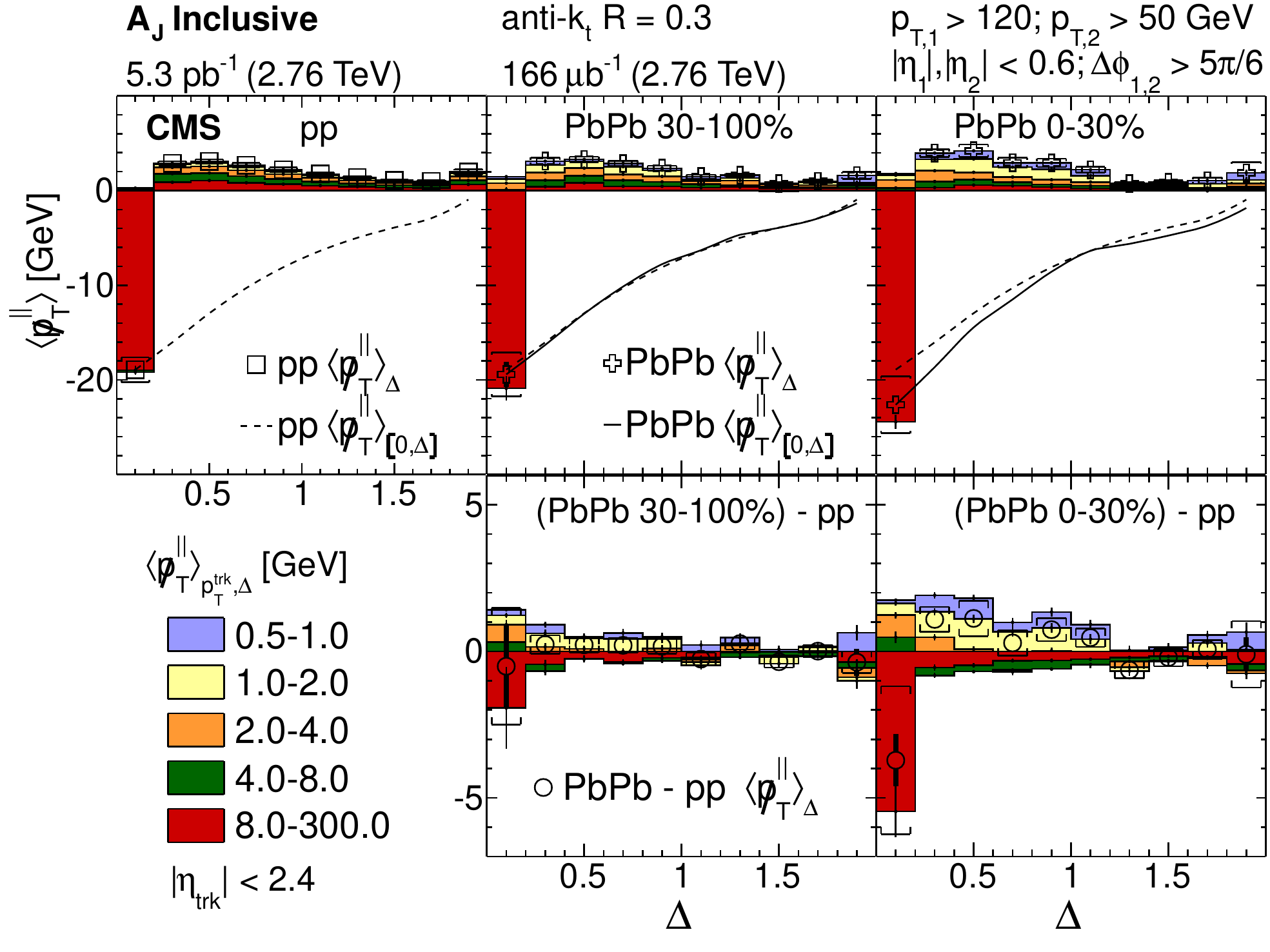}
       \caption{(Color online) Upper row: \MPTpTdR distributions for pp, and for 30--100\% and 0--30\% PbPb data for five
track-$\pt$ ranges (colored boxes), for momentum ranges from $0.5 < \pt < 1\GeV$ (light blue) to $8 < \pt < 300\GeV$ (red), as a function of $\Delta$. Also shown is \MPTdR as a function of $\Delta$ for
pp (open squares) and PbPb data (open plus symbols). Dashed lines (pp) and solid lines (PbPb)
show \MPTcum (\ie integrating the \MPTdR over $\Delta$ from $\Delta =0$ up to the point of interest).
Lower row: Difference between the PbPb and pp \MPTpTdR distributions according to the range in
$\pt$, as a function of $\Delta$ (colored boxes), and difference of \MPTdR as a function of $\Delta$ (open circles), error bars and brackets represent statistical and systematic uncertainties, respectively.}
       \label{fig:Mpt_integrated_Dr}
 \end{figure}

The analysis is performed for pp collisions, and two PbPb centrality selections of
30--100\% and 0--30\%. The resulting differential distributions in \MPTpTdR are shown for different regions of track \pt (in terms of the colored boxes) as a function of $\Delta$ in the upper row of Fig.~\ref{fig:Mpt_integrated_Dr}.
The sum of \MPTpTdR for different $\pt^\text{trk}$ ranges as a function of $\Delta$, \MPTdR, are given by the open markers, and follow the leading jet at small $\Delta$ and subleading jet at large $\Delta$.
The cumulative values, \MPTcum (\ie from summing and smoothing the \MPTdR over bins in $\Delta$, starting at $\Delta = 0$ and ending at the point of interest)
are shown as dashed lines for pp and solid lines for PbPb.
These lines  demonstrate the evolution of the overall \pt balance from small to
large distances relative to the jet axis, reaching an overall balance close to zero only at large radii. The cumulative curve in PbPb collisions for 0--30\% centrality is slightly narrower than for pp collisions.

The distributions in pp collisions have characteristic features, and understanding these is important for interpreting the PbPb results.
The magnitude of the \MPTdR in the first bin, with $\Delta < 0.2$, is related to the average dijet imbalance, and takes a negative value indicating that the momentum projection points along the direction of the leading jet. In the rest of the $\Delta$ bins, \MPTdR takes a positive value, and \MPTpTdR for lower track \pt make up larger fractions of \MPTdR. We refer to the  \MPTpTdR and \MPTdR for bins with $\Delta > 0.2$ as the ``balancing distribution" of the corresponding quantity, because they reduce the large \pt imbalance observed in the first bin in $\Delta$. The balancing distribution has a peak in the range $0.4 < \Delta < 0.6$, which is at the most likely $\Delta$ position for a third jet relative to the subleading jet.

In PbPb collisions, the peak of the balancing $\MPTdR$ distribution shifts towards smaller angles ($0.2 < \Delta < 0.4$). This can be due to the modification in the fragmentation of the leading and subleading jets after quenching, as it occurs at angles close to their axes, where the low-\pt particles make largest contributions. It is therefore not possible to claim a direct relation between the peak position of the balancing $\MPTdR$ distribution and the location of other jets in the event, unless only the highest-\pt particles are considered, \ie not likely to be related to the leading and subleading jets at large $\Delta$ values. The peak position of the balancing \MPTpTdR distribution of the highest-\pt particles is  located at the same place as in pp collisions ($0.4 < \Delta < 0.6$), but with smaller magnitude. This suggests that the position of a third jet relative to the subleading jet is not modified significantly. However, the magnitude of \MPTpTdR for tracks with $8 < \pt < 300\GeV$ associated with the third jet can be reduced for several reasons, such as quenching of the third jet, which makes its fragmentation softer, or a change in the ordering of the jets relative to original partonic conditions, \ie leading parton losing more energy compared to the subleading parton, which causes the third jet to be found in the leading jet hemisphere, instead of the subleading jet hemisphere.

A comparison of pp and PbPb collisions is provided in the lower row of Fig.~\ref{fig:Mpt_integrated_Dr}, showing
the difference in PbPb and pp for \MPTpTdR, and \MPTdR as a function of $\Delta$. For central events, the first bin with $\Delta < 0.2$ \MPTpTdR for high-\pt tracks and \MPTdR point in the leading jet direction, although the excess is not significant. While in the second bin with $0.2 < \Delta < 0.4$, there is a significant positive excess in \MPTdR. The excess towards the subleading jet in this bin may either be because the leading jet is narrower, or the subleading jet wider in PbPb collisions compared to pp collisions. The excess in $\MPTdR$ along the subleading jet direction extends up to larger angles ($\Delta \approx 1$--1.2), with decreasing significance. In this angular range, there is an excess in \MPTpTdR for tracks with \pt that fall in the ranges of 0--0.5, 0.5--1, and 1--2\GeV, and a depletion for particles with $\pt > 4\GeV$.
This is consistent with results shown in the
previous section and earlier CMS studies that demonstrate that the small-angle imbalance towards the leading jet
is compensated by particles of small \pt emitted at large angles to the jet axes~\cite{CMS_dijet2010}.

 \begin{figure}[h!t]
\centering
       \includegraphics[width=0.95\textwidth]{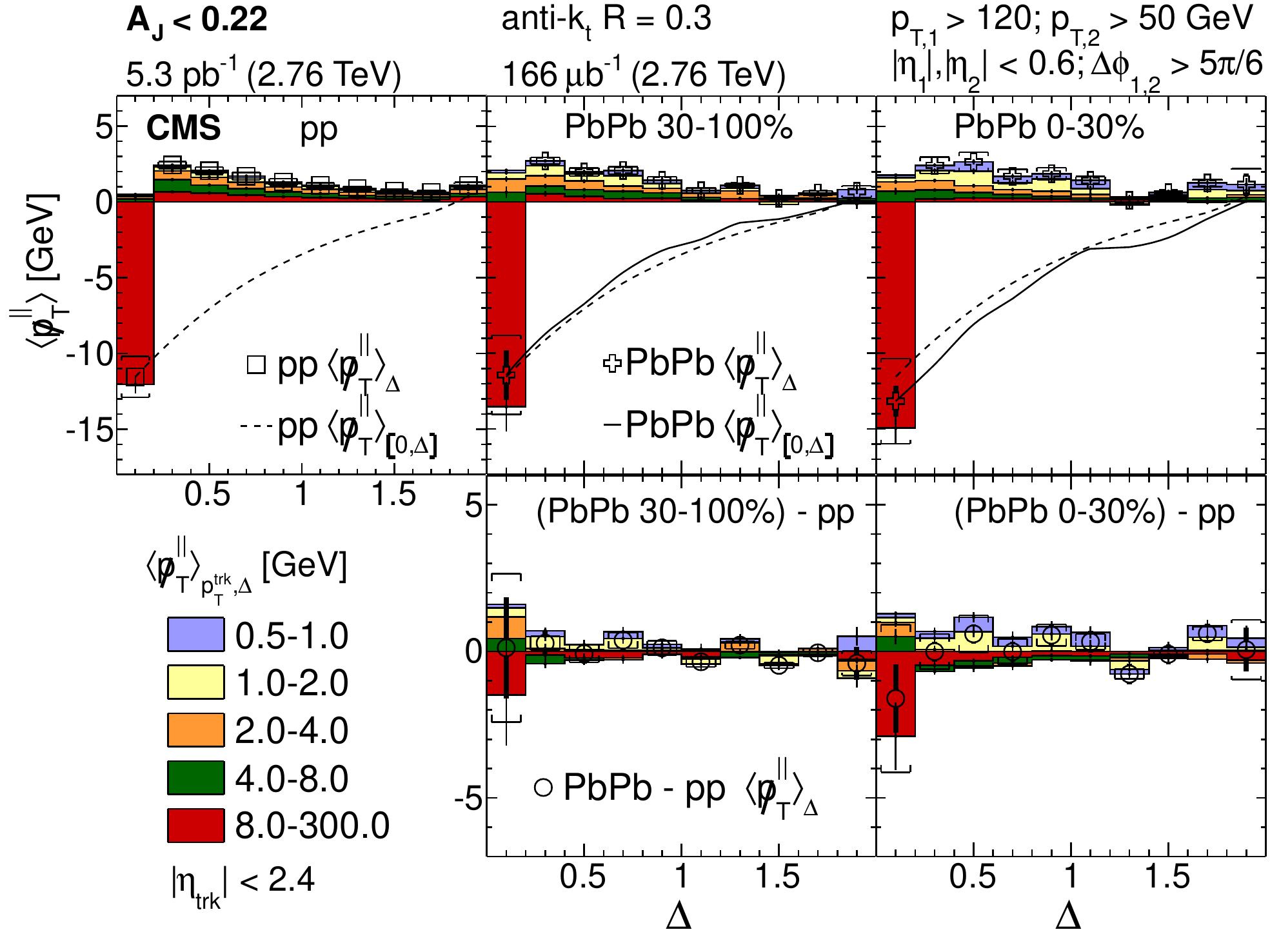}
 \caption{(Color online) Same as Fig.~\ref{fig:Mpt_integrated_Dr}, but with a balanced dijet selection ($\AJ < 0.22$).
 Upper row: \MPTpTdR distributions for pp, and for 30--100\% and 0--30\% PbPb data for five
track \pt ranges (colored boxes), as a function of $\Delta$. Also shown is \MPTdR as a function of $\Delta$ for
pp (open squares) and for PbPb data (open plus symbols). Dashed lines (pp) and solid lines (PbPb)
show \MPTcum (\ie integrating the \MPTdR over $\Delta$ from $\Delta = 0$ up to the point of interest).
Lower row: Difference in the \MPTpTdR distributions for the PbPb and pp according to the range in
$\pt$, as a function of $\Delta$ (colored boxes), and difference of \MPTdR as a function of $\Delta$ (open circles). Error bars and brackets represent statistical and systematic uncertainties, respectively. The y-axis range on the top panels are smaller than in Fig.~\ref{fig:Mpt_integrated_Dr}.
}
\label{fig:Mpt_integrated_Dr_aj0}
 \end{figure}

\subsection{Study of the \texorpdfstring{\pt}{pT} balance in pp and PbPb collisions, as a function of opening angles around jets in bins of \texorpdfstring{$\AJ$}{AJ}}
\label{sec:results_AngDep_AJ}

\begin{figure}[h!t]
\centering
\includegraphics[width=0.95\textwidth]{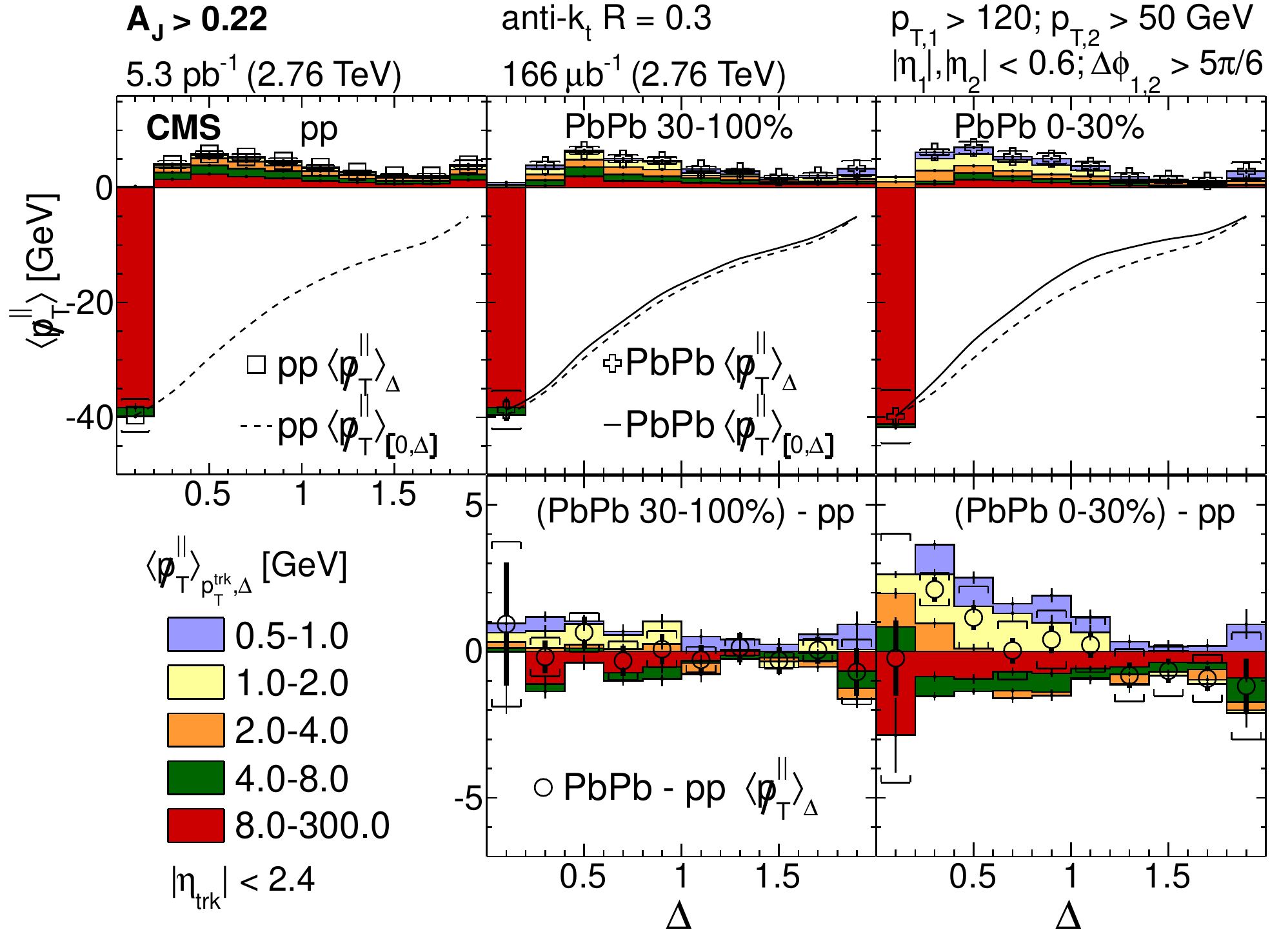}
\caption{(Color online) Same as Fig.~\ref{fig:Mpt_integrated_Dr}, but with an unbalanced dijet selection ($\AJ > 0.22$).
Upper row: \MPTpTdR distributions for pp, and for 30--100\% and 0--30\% PbPb data for five
track $\pt$ ranges, as a function of $\Delta$. Also shown is \MPTdR as a function of $\Delta$ for
pp and for PbPb data. Dashed lines (pp) and solid lines (PbPb)
show \MPTcum (\ie integrating the \MPTdR over $\Delta$ from $\Delta = 0$ up to the point of interest).
Lower row: Difference in the \MPTpTdR distributions for the PbPb and pp. Error bars and brackets represent statistical and systematic uncertainties, respectively. The $y$-axis range on the top panels are larger than in Fig.~\ref{fig:Mpt_integrated_Dr}.
}
       \label{fig:Mpt_integrated_Dr_aj22}
 \end{figure}

More information can be obtained by repeating the previous study as a function of dijet asymmetry
$\AJ$. The results for a sample containing more balanced dijets ($\AJ < 0.22$) is shown in Fig.~\ref{fig:Mpt_integrated_Dr_aj0}, again
comparing pp data with two PbPb centrality bins. As expected, \MPTdR and \MPTpTdR for all track \pt take smaller values compared to inclusive $\AJ$ selection, meaning that events with a more balanced dijet selection show an overall better \pt balance
in both small $\Delta < 0.2$, as well as larger $\Delta$. This is also seen in the difference in \MPTdR for PbPb and pp collisions, although,
as before, an preference of \MPTpTdR for low-$\pt$ tracks to point along the subleading side can be seen for central PbPb events.

Complementary to the selection of more balanced dijets, Fig.~\ref{fig:Mpt_integrated_Dr_aj22} shows a selection for unbalanced dijets
with $\AJ > 0.22$. The $\AJ$ selection is reflected in the overall larger contributions in the small- and large-angle regions relative to the jet axes.
This large $\AJ$ selection, which enhances the fraction of jets having undergone significant energy loss in PbPb collisions, also
enhances the differences between PbPb and pp, as shown in the lower row of Fig.~\ref{fig:Mpt_integrated_Dr_aj22}.

It is important to note that in pp collisions, only 30\% of selected dijet events have $\AJ > 0.22$, but this number increases to 42\% for central PbPb selections. This again suggests the presence of an additional mechanism creating asymmetric dijets in PbPb, \ie parton
energy loss in the medium.
Consistent with this picture, the $\AJ$ dependence of the \MPTpTdR distributions in PbPb and pp collisions and their difference suggests that
asymmetric dijet systems in pp and PbPb collisions are created through different mechanisms, with semi-hard radiation (\eg, three-jet events) dominating
pp collisions. In contrast, a large fraction of asymmetric dijet events in PbPb is created through a differential energy loss mechanism as the partons
traverse the medium, which leads to the observed excess in \MPTpTdR for the low-\pt bins. The depletion of high-\pt particle contributions at large angles in PbPb is more dominant with $\AJ > 0.22$ relative to an inclusive $\AJ$ selection, because of the difference in relative fractions of three-jet events among all selected events.

\subsection{Dependence of dijet asymmetry on \texorpdfstring{\pt}{pT} balance and multiplicity difference in jet hemispheres}

\begin{figure}[h!t]
\centering
     \includegraphics[width=1.0\textwidth]{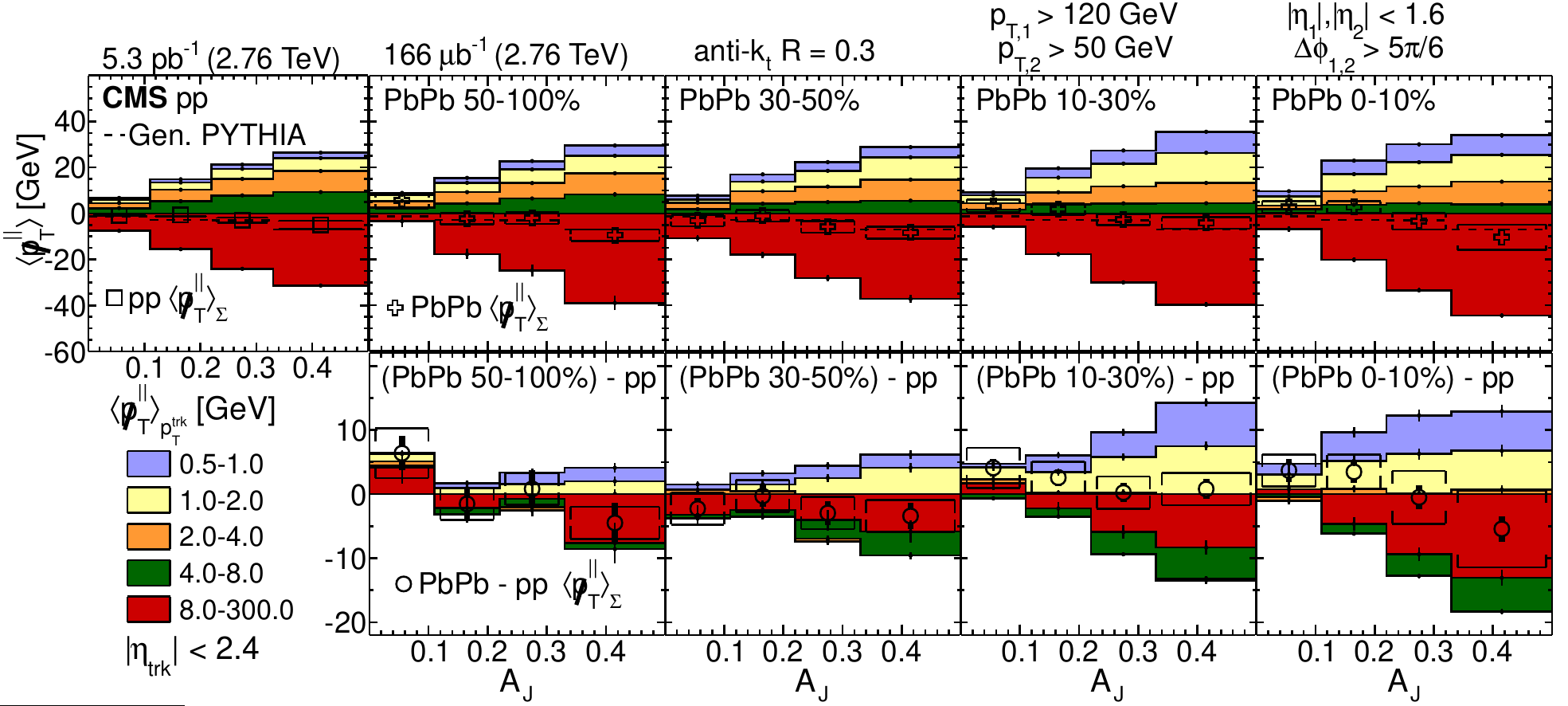}
     \caption{(Color online) Upper row has \MPTpT and \MPTsum in pp collisions (leftmost) and in four selections of PbPb for collision centralities from 50--100\% to 0--10\%. The open markers show \MPTsum, \pt balance for tracks with $0.5 < \pt < 300\GeV$, while the colored boxes show the \MPTpT contributions for different track \pt ranges.
For each panel, \MPTpT and \MPTsum values are shown as a function of dijet asymmetry.
The lower row shows the difference  between \MPTpT and \MPTsum for PbPb and pp data. Error bars and brackets represent statistical and systematic uncertainties, respectively.}
     \label{fig:lPtPlot}
 \end{figure}

To study the \pt flow relative to the dijet system as a function of event properties, such as centrality and $\AJ$, in more detail, the\MPTpTdR is summed over all annuli to obtain \MPTpT, \ie the average \pt balance in the event calculated for a given range of track $\pt$.
In Fig.~\ref{fig:lPtPlot}, we display \MPTpT for different ranges of track \pt (displayed in terms of the colored boxes) as a function of $\AJ$, ranging from almost balanced to very unbalanced dijets in pp collisions, and in four selections of PbPb centrality from most peripheral to most central. The balance in the event for all tracks with $\pt > 0.5\GeV$, denoted as \MPTsum, which is obtained by adding up the \MPTpT for different $\pt$ ranges, is also included, and shown as open markers, with associated systematic uncertainties as brackets around the points. In PbPb events, overall \pt is balanced to better than $10\GeV$, \ie $|\MPTsum|<10\GeV$ for all $\AJ$ selections. The small negative trend in \MPTsum as a function of $\AJ$ is observed also in pp events, and in generator-level \PYTHIA events, once the \pt threshold set on charged particles and the acceptance of the tracker are imposed.

When selecting events containing dijets with $\AJ > 0.11$, an expected excess of high-$\pt$ particles in the direction
of the leading jet (indicated by the red areas in Fig.~\ref{fig:lPtPlot}) is seen for all selections in pp and PbPb collisions.
For pp and peripheral PbPb collisions, this excess is mostly balanced
by particles with intermediate \pt of 2--8\GeV. Going
to more central collisions, \MPTpT on the subleading jet side is modified from the intermediate $\pt$ range towards low $\pt$ (0.5--2\GeV). This effect is most pronounced for events with large $\AJ$ in central PbPb collisions.

The lower row of Fig.~\ref{fig:lPtPlot} shows the difference between \MPTpT in PbPb and pp collisions, after requiring the specific  PbPb collision centralities and dijet imbalance. While the
contributions from different $\pt$ ranges are similar for pp and peripheral PbPb collisions, a difference can be seen
for central collisions, where a significant excess of low-$\pt$ charged particles is observed for asymmetric jets in PbPb collisions. Systematic uncertainties are shown only for \MPTsum, and not for \MPTpT. Uncertainties in \MPTsum provide an upper bound on systematic uncertainties for individual \pt ranges, as uncertainties in low-\pt particles are, in fact, significantly smaller. The excess observed in low-\pt particles in the range of 0.5--2\GeV has therefore a significance of 3--4 standard deviations for $\AJ > 0.11$ for most central events. The difference in $\langle \PTslash^{\parallel} \rangle$ between PbPb and pp collisions for all tracks with $\pt > 0.5\GeV$ is consistent with zero across all centrality and $\AJ$ selections.

The overall \pt balance observed through \MPTsum in PbPb events agrees with pp events, within systematic and statistical uncertainties, over all ranges of $\AJ$ and centrality, while the \MPTpT distributions show excess of low-\pt particles. This implies that there are more particles in the subleading jet hemispheres compared to the leading jet hemispheres, because more particles are required to obtain the same \pt sum.

\begin{figure}[h!t]
\centering
\includegraphics[width=\textwidth]{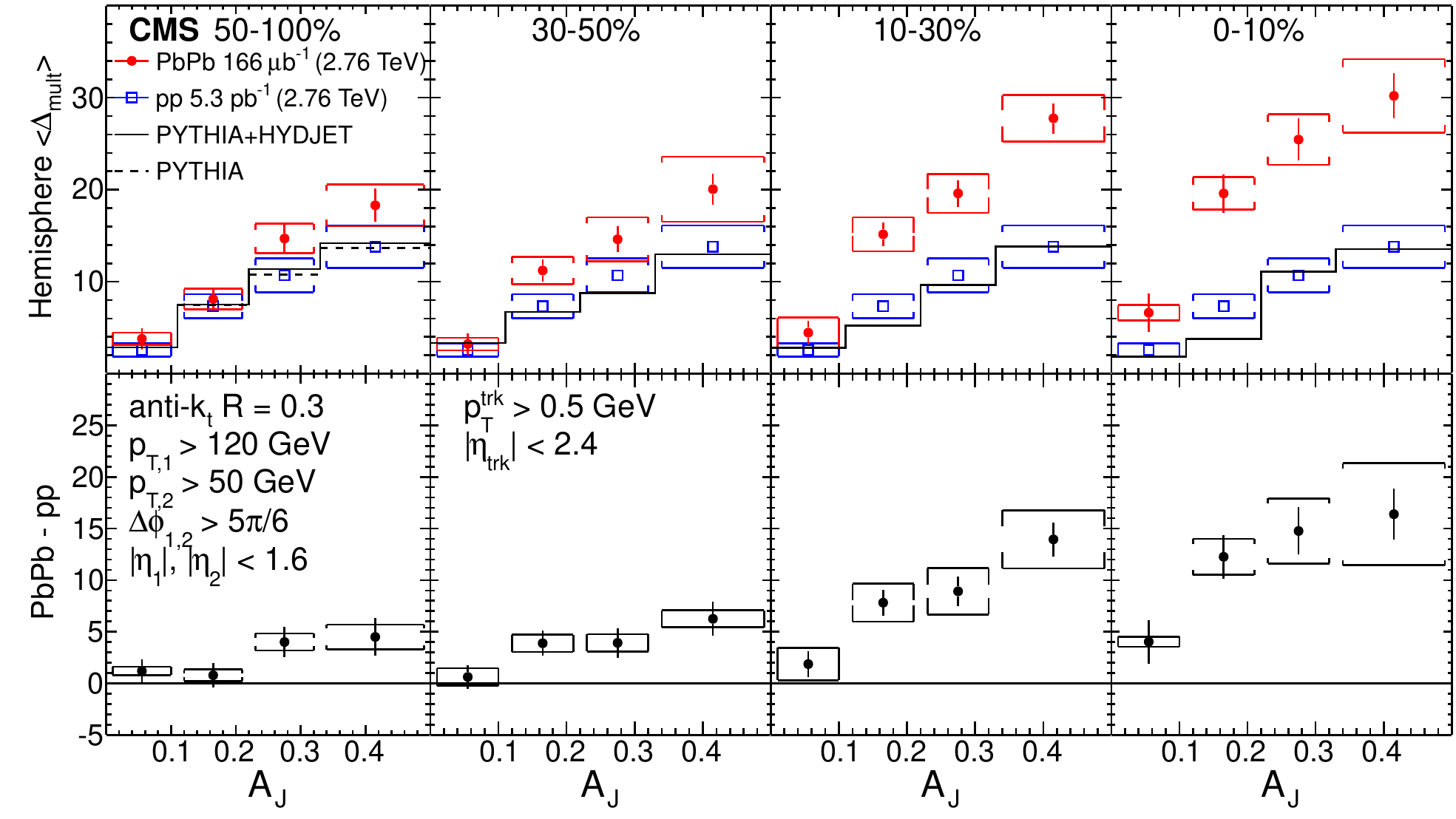}
\caption{(Color online) Upper panels show the comparison of the mean difference in multiplicity $\langle\Delta_\text{mult}\rangle$ between the subleading jet hemisphere and
leading jet hemisphere, as a function of dijet asymmetry $\AJ$ for pp (blue squares), PbPb (red filed circles), \PYTHIA (dashed histogram),
and \textsc{pythia+hydjet} events (black histogram). The centralities of PbPb collisions are 50--100\%, 30--50\%, 10--30 \%, and 0--10\%, respectively, from leftmost to rightmost panel. Lower panels provide the difference in $\langle\Delta_\text{mult}\rangle$ between PbPb and pp collisions. Statistical and systematic uncertainties are shown as error bars and brackets, respectively.}
\label{fig:Result_MultiplicityDifference_AJ}
\end{figure}

Figure~\ref{fig:Result_MultiplicityDifference_AJ} shows the mean difference in multiplicities between leading and subleading jet hemispheres, denoted as $\langle \Delta_\text{mult} \rangle$,
as a function of $\AJ$ and collision centrality. The $\langle \Delta_\text{mult} \rangle$ is presented
for both PbPb and pp collisions. Measurements in pp collisions are
in good agreement with \PYTHIA and \textsc{pythia+hydjet} simulations. In general, the $\langle\Delta_\text{mult}\rangle$
increases as a function of $\AJ$ in pp, PbPb, \PYTHIA, and \textsc{pythia+hydjet} events.
The events in pp collisions with large $\AJ$ contain a larger fraction of three-jet or multijet events, where more particles are produced in the direction of the subleading jet.
The observed increase in $\langle\Delta_\text{mult}\rangle$ for pp collisions with increasing $\AJ$ is therefore expected.
Going from peripheral  (50--100\%) to central (0--10\%) PbPb events,  for
a given $\AJ$ selection an excess
in $\langle\Delta_\text{mult}\rangle$ is visible compared to pp collisions.
The difference in $\langle \Delta_\text{mult} \rangle$ between pp and PbPb collisions increases monotonically
as a function of $\AJ$ at all collision centralities, with the biggest effect seen for most
central PbPb collisions. This is consistent with the expected dependence of medium-induced energy loss
on collision centrality, where systems of the largest size (\ie smallest centrality) should
show the largest medium-related effects.
The multiplicity difference is up to ${\approx}15$ particles in the most central  0--10 \% collisions.

\subsection{Dependence of transverse momentum balance on jet distance parameter \texorpdfstring{$R$}{R}}
\label{sec:results_RDep}

In pp collisions, jets clustered with small $R$ are narrower and fragment into components with higher \pt than jets clustered with large $R$. In addition, using small $R$ tends to bias the clustered jets to contain a larger fraction of quark jets~\cite{Dasgupta:2014yra,Cacciari:2008gd}. Changing the $R$ parameter can provide a handle on the size and shower profiles of individual jets. In heavy ion collisions, studying the $R$ dependence of momentum flow in dijet events makes it possible to investigate whether jet quenching mechanisms act differently on jets with different fragmentation patterns on a jet-by-jet basis.

It is important to note that there is an overlap in the final set of dijet events obtained for different $R$ parameters, and therefore it is not possible to interpret the dependence of the \pt-balance distributions on $R$ as simply a dependence on jet size. A change in $R$ can induce a modification in $\PTslash^{\parallel}$ in two ways: events that satisfy the dijet requirements for one $R$ can fail for another $R$ value, or events that satisfy the dijet requirements for both $R$ parameters, but for which the ordering of jets change, can impact $\phi_{\text{dijet}}$, as well as the value of parameters used in the binning of the measurements, such as $\AJ$ and $\Delta$.

The requirements on the \pt of leading and subleading jets are the main sources of variations in the final set of dijet events for different $R$ parameters. For each $R$, a jet \pt selection translates into a different requirement on initial parton \pt. A smaller fraction of the initial energy of the parton is recovered using jets of smaller size. Although fewer events pass the dijet requirement for $R = 0.2$ jets, strictly speaking, such events do not form a subset of dijet events with larger $R$ parameters. A small fraction of $R = 0.2$ dijet events (4--7\% in PbPb collisions and 2--4\% in pp collisions) does not satisfy the dijet requirements for other $R$ values, mainly because jets fall outside of the $\eta$ range or the $\Delta\phi$ requirement for the dijet pair. This can happen because of the merging of the subleading and third jets, and because of the resolution in jet angular direction. Such events make up a statistically negligible contribution to the results and are therefore not the focus of the discussion.

\begin{table}[h!t]
\centering
\topcaption{
Overlap in event selections for 0--100\% PbPb and pp collisions. The second column gives the percentage of events that pass dijet selections and a tight pseudorapidity requirement ( $\abs{\eta}<0.6$ ) for $R = 0.5$, and an additional dijet selection also required for a smaller $R$ value. In columns 3--6 the leading and subleading jets with $R = 0.5$ are matched to the leading and subleading jets with smaller $R$ values, requiring only $R = 0.5$ selection on jets. The third column shows the percentage of these events where both leading and subleading jets point in the same direction ($\Delta_{i} = \sqrt{ \smash[b]{ (\eta_{i}^{R}-\eta_{i}^{R = 0.5})^{2} + (\phi_{i}^{R}-\phi_{i}^{R = 0.5})^{2}}} < 0.5$ for $i=1$ and $2$). The average value of the ratio of \pt of the leading and subleading jets at jet for a given $R$, to their \pt for $R = 0.5$ are shown in the fourth and fifth columns, respectively. The sixth column shows percentage of events in which subleading jets with the given R parameter match the $R = 0.5$ leading jet, and the leading jet matches the $R = 0.5$ subleading jet.
}
\label{table_compare_R5}
\cmsTableResize{
\begin{tabular}{cy{5}y{5}y{5}y{5}y{5}}
\hline
& \multicolumn{1}{c}{Additional}    &  \multicolumn{1}{c}{Matched} & & & \multicolumn{1}{c}{Swapped}  \\
$R$   & \multicolumn{1}{c}{dijet selection [\%] }   & \multicolumn{1}{c}{ jet directions [\%] } & \multicolumn{1}{c}{ $\langle p_{\rm T,1}^{R}/p_{\rm T,1}^{R=0.5} \rangle$ } & \multicolumn{1}{c}{$\langle  p_{\rm T,2}^{R}/p_{\rm T,2}^{R=0.5}\rangle$ }  & \multicolumn{1}{c}{ jet directions [\%] } \\
\hline
\multicolumn{6}{c}{PbPb}\\
\hline
0.2 & 48 , 2 & 83 , 5 & 0.89 , 0.001 & 0.79 , 0.002 & 10 , 3\\
0.3 & 62 , 2 & 90 , 4 & 0.93 , 0.002 & 0.88 , 0.004 & 7  , 3\\
0.4 & 77 , 1 & 94 , 3 & 0.96 , 0.002 & 0.94 , 0.005 & 3 , 2\\
\hline
\multicolumn{6}{c}{pp}\\
\hline
0.2 & 58 , 2 & 83 , 5 & 0.91 , 0.001 & 0.83 , 0.002 & 14 , 3 \\
0.3 & 73 , 2 & 90 , 4 & 0.95 , 0.001 & 0.90 , 0.001 & 8 , 3 \\
0.4 & 86 , 1 & 95 , 3 & 0.98 , 0.001 & 0.96 , 0.001 & 4 , 2 \\
\hline
\end{tabular}}
\end{table}

The fraction of events that pass the dijet selection both for the largest $R = 0.5$ and for other values are shown in the second column of Table~\ref{table_compare_R5},
without matching the directions of the jets. Compared to pp collisions, the fraction of events that pass both cutoffs on jets is reduced in PbPb collisions more rapidly as $R$ decreases. This observation is qualitatively consistent with the measurement showing that inclusive jet suppression is smaller in PbPb collisions for large $R$ values~\cite{Aad:2012vca}, which can be interpreted as due to the recovery of part of the energy lost in the initial hard scatter of partons.

Additional information can therefore be extracted by requiring the leading and subleading jets with a given $R$ to be in the same direction as the corresponding jets found using $R = 0.5$. As shown in the third column of Table~\ref{table_compare_R5}, the fraction of such events is similar for pp and PbPb collisions. These events produce almost no change in $\phi_{\rm{dijet}}$ and the jet axes, which change only slightly due to jet angular resolution, and therefore yield approximately the same $\PTslash^{\parallel}$.
However, these events can accommodate the change in the \pt of jets that originate from the same initial hard-scattered parton for different $R$ parameters. For jets matched to each other spatially, the ratio of the \pt of the leading or subleading jet at some given $R$ to respective jets with $R = 0.5$, $\langle p_{\rm T,1(2)}^{R}/p_{\rm T,1(2)}^{R = 0.5} \rangle$, is calculated and the values are shown in columns 4 and 5 in Table~\ref{table_compare_R5}. As expected, in both PbPb and pp collisions, $\langle p_{\rm T,1}^{R}/p_{\rm T,1}^{R = 0.5} \rangle$ and $\langle p_{\rm T,2}^{R}/p_{\rm T,2}^{R = 0.5} \rangle$ are reduced as $R$ gets smaller. In PbPb collisions, a smaller fraction of jet \pt is recovered at small $R$ for both the leading and subleading jets, which may be due to the broadening of quenched jets. This effect is larger for the subleading than for the leading jet.

As $R$ parameters become smaller, leading and subleading jets fall below the \pt requirements. Most of the time, the leading jet satisfies the \pt selection for $R = 0.5$, but falls below the threshold for smaller $R$, because the subleading jet \pt is already biased towards values above the 50\GeV threshold by the leading jet with $\pt > 120\GeV$ in the event. However, as shown in Figs.~\ref{fig:Mpt_integrated_Dr_aj0} and~\ref{fig:Mpt_integrated_Dr_aj22}, for $R = 0.3$ jets the \MPTpTdR signal is dominated by dijet events with large imbalance, which is true for all other $R$ parameters as well. For events with $\AJ > 0.22$, $\langle p_{\rm T,2} \rangle \approx$ 70--80\GeV is sufficiently close to the 50\GeV threshold for subleading jets falling below the threshold to create sizable effects on the results.

The last column of Table~\ref{table_compare_R5} gives the fraction of events with swapped leading and subleading jets compared to those with $R = 0.5$. For these events, the $ \PTslash^{\parallel}$ has an opposite sign relative to the value for $R = 0.5$, as $\phi_{\rm{dijet}}$ points in the opposite hemisphere.
Especially in pp collisions, swapping of the leading and subleading jet is the main source of events in which the jet directions are not matched. In PbPb collisions, swapping is slightly less frequent than in pp collisions, suggesting that the third jet may be replacing the subleading jet. For events that satisfy dijet requirements for different R parameters, the $\PTslash^{\parallel}$ in each event can still change as a function of $R$ because of the swapping of jets in the dijet pairs, and the replacement of the subleading jet by the third jet.

\begin{figure}[h!t]
\centering
\includegraphics[width=0.95\textwidth]{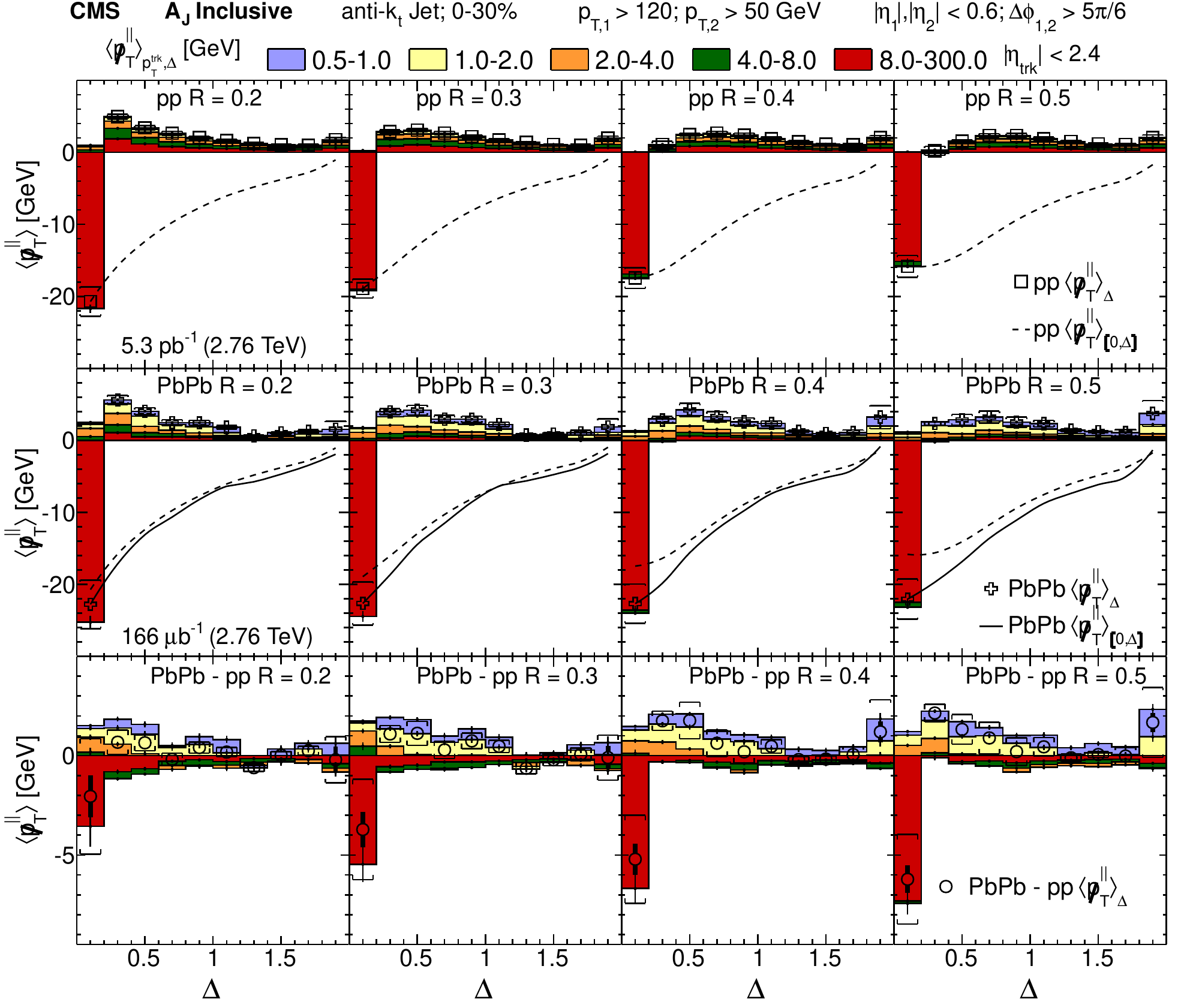}
\caption{ (Color online) Upper row shows \MPTpTdR in pp collisions as a function of $\Delta$, for a distance parameter $R = 0.2$, $0.3$, $0.4$, and $0.5$, from left to right for different ranges of track $\pt$, and \MPTdR (\ie \MPTpTdR summed over all $\pt$  for a given $\Delta$ bin).
Dashed lines indicate cumulative results for \MPTcum in pp, for each distance parameter (\ie integrating \MPTdR over the $\Delta$ range from $\Delta$ = 0 to the point of interest).
Middle row provides \MPTpTdR and \MPTdR in PbPb collisions of centrality range 0--30\% as a function of $\Delta$, for distance parameters $R = 0.2$, $0.3$, $0.4$, and $0.5$ from left to right. Solid line indicates \MPTcum in PbPb for each distance parameter.
Lower row has the difference between PbPb and pp. Error bars and brackets represent statistical and systematic uncertainties, respectively. The results are inclusive in the dijet asymmetry parameter $\AJ$.}
\label{fig:Mpt_delR_incAj_RDep030}
 \end{figure}

The dependence of \MPTpTdR on $\Delta$ and $R$ is shown in Fig.~\ref{fig:Mpt_delR_incAj_RDep030}, without any $\AJ$ requirement, for pp and for PbPb events with 0--30\% centralities. The $R$-dependent evolution in pp collisions, which is attributed to the softening and broadening of jets, can be seen as a shift in the position of the sign change of \MPTpTdR and as a decrease in the total imbalance within the jet cones \mbox{ $\Delta \lesssim 0.2$--0.4 }. Moreover, the peaking point of the balancing distribution shifts towards larger $\Delta$, as jet distance parameter $R$ increases (from $\Delta =$ 0.2--0.4 for $R = 0.2$ jets, to $\Delta =0.6$--1.0 for $R = 0.5$ jets). As stated for $R = 0.3$ jets in Section~\ref{sec:results_AngDep}, the peak position is correlated with the most likely position of the third jet relative to the subleading jet, which also moves to larger angles by increasing $R$.

In the PbPb system, the peak also shifts towards greater $\Delta$, but less than in pp collisions due to the additional soft particles at small angles associated to the quenching of the dijet pair and reduction in the number of high-\pt particles associated with the third jet. In the PbPb$-$pp bottom panels, this manifests in the depletion of higher ranges at \pt, 4--8 and 8--300\GeV, which shift to greater angular distance with increasing $R$. There is a modest increase observed in the excess in the \pt ranges of 0.5--1 and 1--2\GeV with increasing $R$. The overall distribution in the low-\pt excess in PbPb relative to pp does not change significantly with the distance parameter, and especially not at larger angular distance $\Delta$.

There is a hint that the \MPTcum distribution in central PbPb collisions, shown by the black curves in Fig.~\ref{fig:Mpt_delR_incAj_RDep030}, is narrower than in pp collisions, shown by the dashed black curves, meaning that the slope is larger in PbPb relative to pp collisions. This becomes slightly more significant at $R = 0.5$, where bias in gluon or quark jets that have large angular width becomes smaller. This is also reflected in the increase in the magnitude of \MPTdR in the leading jet direction in the first bin, and in the subleading jet direction in the second bin. This modification is dominated by particles with $\pt > 2\GeV$, and may arise from quenching effects, causing leading jets to narrow or subleading jets to widen in central PbPb relative to pp collisions.

\begin{figure}[h!t]
    \centering
       \includegraphics[width=0.95\textwidth]{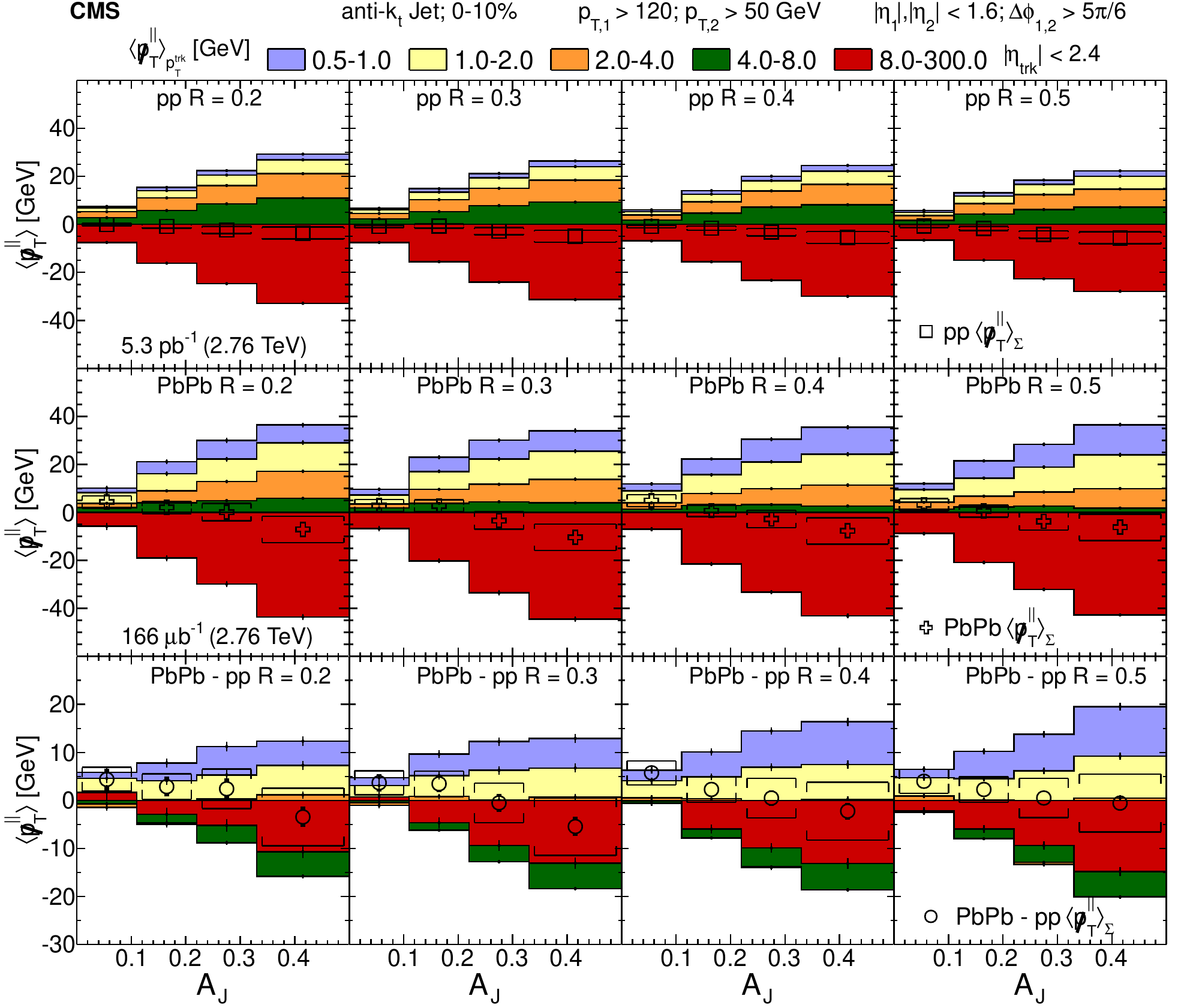}
\caption{(Color online) Upper row shows \MPTpT (the individual track \pt) and \MPTsum (sum over all ranges of track \pt) as a function of $\AJ$ in pp collisions for distance parameters $R = 0.2$, $0.3$, $0.4$, and $0.5$, from left to right. The dijet asymmetry ranges from almost balanced ($\AJ < 0.11$) to unbalanced ($\AJ > 0.33$) dijets. Middle row provides \MPTpT and \MPTsum as a function of $\AJ$ in PbPb collisions of centrality range 0--10\%, for distance parameter $R = 0.2$, $0.3$, $0.4$, and $0.5$, from left to right. Lower row has the difference PbPb $-$ pp of the \MPTpT, and \MPTsum, which are shown in the upper panels. Error bars and brackets represent statistical and systematic uncertainties, respectively.}
\label{fig:Mpt_aj_RDep010}
 \end{figure}

To summarize the dependence of differences in \pt balance among different $R$ bins on $\AJ$, and to investigate the observed changes in the associated track \pt spectrum in more central events, our measurement of the dependence of the \pt balance on $R$ and $\AJ$, is shown in Fig.~\ref{fig:Mpt_aj_RDep010} for pp and 0--10\% central PbPb events, respectively, in the top and middle rows. The leftmost panels correspond to a selection of $R=0.2$ jets, while the rightmost panels correspond to $R=0.5$. For pp collisions, there is a slight decrease in the magnitude of signal in each \pt range as $R$ increases. This behavior is consistent with the observed reduction in the incone \MPTpTdR for high-\pt tracks with $\Delta < 0.2$ shown in the top panels of Fig.~\ref{fig:Mpt_delR_incAj_RDep030} as a function of $R$, which was discussed above, and is also observed in generator-level \PYTHIA. This kind of behavior is not observed in central PbPb events.

The bottom row of Fig.~\ref{fig:Mpt_aj_RDep010} displays the difference between PbPb and pp results. The $R$ parameter is correlated with a small change in the magnitude of the \MPTpT excess of low-\pt particles, as jets of larger $R$ give a greater excess. When \pt ranges 0.5--2.0\GeV are combined, the increase in the low-\pt excess becomes more significant. The systematic uncertainties shown in the plot are dominated primarily by the \pt range 8.0--300.0\GeV, and as such cannot be used to characterize the significance of \MPTpT in the low track-\pt ranges, nor the slight dependence on the distance parameter in the low-\pt excess. The sum of track \pt ranges \MPTsum is insensitive to the distance parameter, and the difference between PbPb and pp collisions is consistent with zero for all $R$ values.

\begin{figure}[h!t]
\centering
\includegraphics[width=\textwidth]{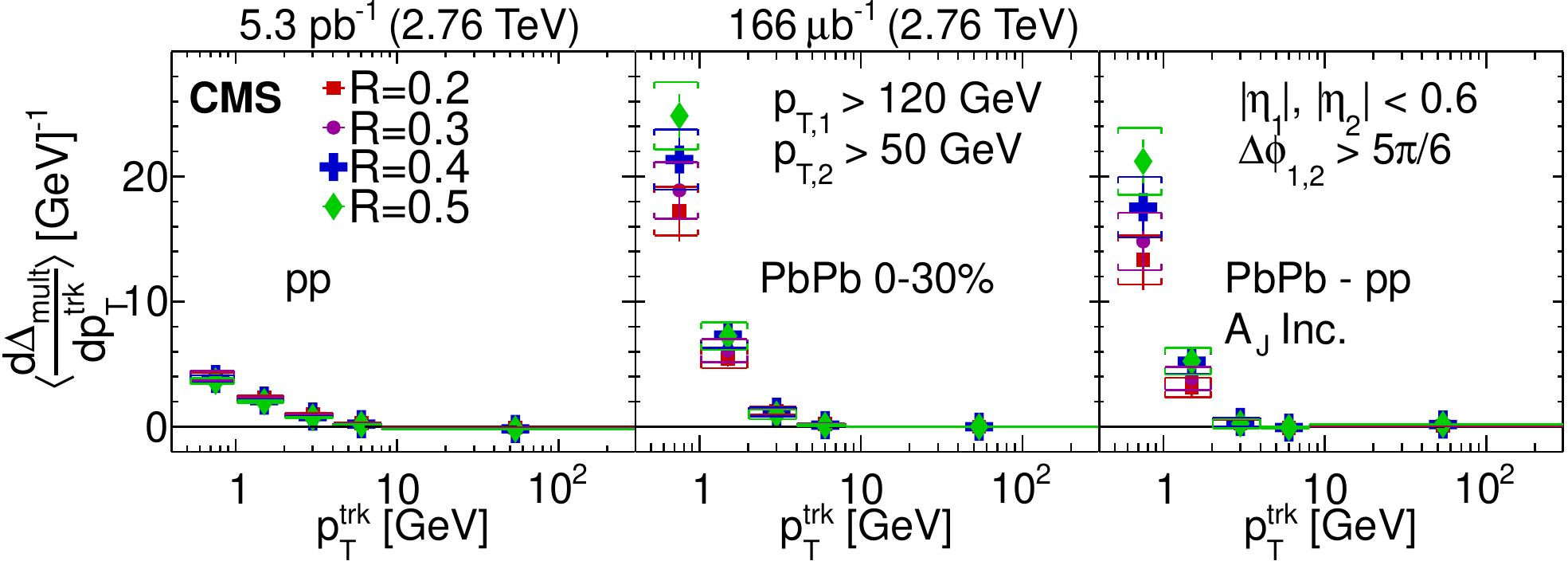}
\caption{(Color online) Difference in differential multiplicity $\langle {\rd\Delta_\text{mult}}/{\rd p_{\mathrm{T}^\text{trk}}}\rangle$ between the away-side and leading-jet hemispheres as a function of track $\pt$, using an inclusive dijet asymmetry selection. Left panel has measurements in pp for jet radii $R = 0.2$, $0.3$, $0.4$, and $0.5$, and the middle panel displays similar measurements in PbPb. Right panel provides the difference in $\langle{\rd \Delta_\text{mult}}/{\rd \pt^\text{trk}}\rangle$ between PbPb and pp collisions for each momentum range. Systematic uncertainties are shown as boxes. Error bars represent statistical uncertainties.}
\label{fig:Result_dNdPT_AJInc}
\end{figure}

Finally, the multiplicity associated with excess of low-\pt particles shown in Figs.~\ref{fig:Mpt_delR_incAj_RDep030} and~\ref{fig:Mpt_aj_RDep010}, and the charged-particle spectrum for $\langle\Delta_\text{mult}\rangle$ are given in Fig.~\ref{fig:Result_dNdPT_AJInc} for events with 0--30\% centrality, without any $\AJ$ requirement, for several distance parameters in pp and PbPb collisions, and for their difference.

In pp collisions the fragmentation of leading jets with high \pt provides more high-\pt and fewer low-\pt particles in the hemisphere of the leading jet relative to the subleading-jet hemispheres. As a result, $\langle \rd\Delta_\text{mult} / \rd\pt \rangle$ has a positive value for charged particles with $\pt < 8\GeV$ and a negative value for charged particles with $\pt > 8\GeV$. Also, in PbPb collisions, $\langle \rd\Delta_\text{mult} / \rd\pt \rangle$  is positive for particles with $\pt < 8\GeV$ and becomes negative in the last bin, although the spectrum is much steeper, and has a large excess of soft particles. By taking the difference in $\langle \rd\Delta_\text{mult} / \rd\pt \rangle$ between PbPb and pp collisions, a significant excess (${>}5$ standard deviations) is observed at $\pt < 2\GeV$, and a depletion at $\pt > 4\GeV$, while there is only a slight excess in the range $2 < \pt < 4\GeV$. Changing $R$ does not have an effect on the results in pp collisions, while in PbPb collisions there is a small enhancement in the excess for low-\pt charged particles as $R$ is increased from 0.2 to 0.5.

\section{Summary and conclusions}

The transverse momentum flow relative to the dijet axis in PbPb and pp collisions containing jets with large $\pt$ has been studied
using data corresponding to integrated luminosities of 166\mubinv and 5.3\pbinv,
respectively, collected at a nucleon-nucleon center-of-mass energy of 2.76\TeV.
Dijet events were selected containing a leading jet with
transverse momentum $p_{\rm T,1} > 120\GeV$  and a subleading jet with $p_{\rm T,2} > 50\GeV$,
reconstructed using the anti-$\kt$ algorithm, with distance parameters of $R = 0.2$, $0.3$, $0.4$ and $0.5$.
For PbPb collisions, the dijet events show a larger asymmetry in \pt between the leading and subleading
jets than in pp collisions.
The multiplicity, angular, and \pt spectra of the radiation balancing this asymmetry are characterized
using several techniques as a function of PbPb collision centrality and \pt asymmetry.
For a given dijet asymmetry, the imbalance in \pt in PbPb collisions is found to be compensated
by particles at $\pt =$ 0.5--2\GeV, whereas in pp collisions most
of the momentum balance is found in the $\pt$ range of 2--8\GeV, reflecting a softening
of the radiation responsible for the imbalance in \pt of the asymmetric dijet system in PbPb interactions. Correspondingly, a larger
multiplicity of associated particles is seen in PbPb than in pp collisions. Both measurements show larger differences between PbPb and pp  for more central PbPb collisions. The current data provide
the first detailed study of the angular dependence of charged particle contributions to the asymmetry up to large
angles from the jet axis ($\Delta = 1.8$).
Despite the large shift in the \pt spectrum of particles, the angular pattern of energy flow in PbPb events as a function of $\Delta$ matches that seen in pp collisions, especially for small $R$ parameters. The results suggest that either the leading jet is getting narrower, or the subleading jet is getting broader after quenching. In pp collisions, the balancing distribution shifts to larger $\Delta$ with increasing distance parameter R, likely because of the presence of a third jet further away from the dijet axis. The shift is more pronounced than in PbPb collisions, where there is an excess of low \pt particles close to the jet axes.
These results constrain the redistribution of transverse momentum in the modelling of QCD energy loss processes of partons traversing the hot and dense medium created in heavy-ion collisions.

\section*{Acknowledgments}

We congratulate our colleagues in the CERN accelerator departments for the excellent performance
of the LHC and thank the technical and administrative staffs at CERN and at other CMS
institutes for their contributions to the success of the CMS effort. In addition, we gratefully
acknowledge the computing centres and personnel of the Worldwide LHC Computing Grid
for delivering so effectively the computing infrastructure essential to our analyses. Finally, we
acknowledge the enduring support for the construction and operation of the LHC and the CMS
detector provided by the following funding agencies: BMWFW and FWF (Austria); FNRS and
FWO (Belgium); CNPq, CAPES, FAPERJ, and FAPESP (Brazil); MES (Bulgaria); CERN; CAS,
MoST, and NSFC (China); COLCIENCIAS (Colombia); MSES and CSF (Croatia); RPF (Cyprus);
MoER, ERC IUT and ERDF (Estonia); Academy of Finland, MEC, and HIP (Finland); CEA and
CNRS/IN2P3 (France); BMBF, DFG, and HGF (Germany); GSRT (Greece); OTKA and NIH
(Hungary); DAE and DST (India); IPM (Iran); SFI (Ireland); INFN (Italy); MSIP and NRF (Republic
of Korea); LAS (Lithuania); MOE and UM (Malaysia); CINVESTAV, CONACYT, SEP,
and UASLP-FAI (Mexico); MBIE (New Zealand); PAEC (Pakistan); MSHE and NSC (Poland);
FCT (Portugal); JINR (Dubna); MON, RosAtom, RAS and RFBR (Russia); MESTD (Serbia);
SEIDI and CPAN (Spain); Swiss Funding Agencies (Switzerland); MST (Taipei); ThEPCenter,
IPST, STAR and NSTDA (Thailand); TUBITAK and TAEK (Turkey); NASU and SFFR (Ukraine);
STFC (United Kingdom); DOE and NSF (USA).
\hyphenation{Bundes-ministerium Forschungs-gemeinschaft Forschungs-zentren} We congratulate our colleagues in the CERN accelerator departments for the excellent performance of the LHC and thank the technical and administrative staffs at CERN and at other CMS institutes for their contributions to the success of the CMS effort. In addition, we gratefully acknowledge the computing centres and personnel of the Worldwide LHC Computing Grid for delivering so effectively the computing infrastructure essential to our analyses. Finally, we acknowledge the enduring support for the construction and operation of the LHC and the CMS detector provided by the following funding agencies: the Austrian Federal Ministry of Science, Research and Economy and the Austrian Science Fund; the Belgian Fonds de la Recherche Scientifique, and Fonds voor Wetenschappelijk Onderzoek; the Brazilian Funding Agencies (CNPq, CAPES, FAPERJ, and FAPESP); the Bulgarian Ministry of Education and Science; CERN; the Chinese Academy of Sciences, Ministry of Science and Technology, and National Natural Science Foundation of China; the Colombian Funding Agency (COLCIENCIAS); the Croatian Ministry of Science, Education and Sport, and the Croatian Science Foundation; the Research Promotion Foundation, Cyprus; the Ministry of Education and Research, Estonian Research Council via IUT23-4 and IUT23-6 and European Regional Development Fund, Estonia; the Academy of Finland, Finnish Ministry of Education and Culture, and Helsinki Institute of Physics; the Institut National de Physique Nucl\'eaire et de Physique des Particules~/~CNRS, and Commissariat \`a l'\'Energie Atomique et aux \'Energies Alternatives~/~CEA, France; the Bundesministerium f\"ur Bildung und Forschung, Deutsche Forschungsgemeinschaft, and Helmholtz-Gemeinschaft Deutscher Forschungszentren, Germany; the General Secretariat for Research and Technology, Greece; the National Scientific Research Foundation, and National Innovation Office, Hungary; the Department of Atomic Energy and the Department of Science and Technology, India; the Institute for Studies in Theoretical Physics and Mathematics, Iran; the Science Foundation, Ireland; the Istituto Nazionale di Fisica Nucleare, Italy; the Ministry of Science, ICT and Future Planning, and National Research Foundation (NRF), Republic of Korea; the Lithuanian Academy of Sciences; the Ministry of Education, and University of Malaya (Malaysia); the Mexican Funding Agencies (CINVESTAV, CONACYT, SEP, and UASLP-FAI); the Ministry of Business, Innovation and Employment, New Zealand; the Pakistan Atomic Energy Commission; the Ministry of Science and Higher Education and the National Science Centre, Poland; the Funda\c{c}\~ao para a Ci\^encia e a Tecnologia, Portugal; JINR, Dubna; the Ministry of Education and Science of the Russian Federation, the Federal Agency of Atomic Energy of the Russian Federation, Russian Academy of Sciences, and the Russian Foundation for Basic Research; the Ministry of Education, Science and Technological Development of Serbia; the Secretar\'{\i}a de Estado de Investigaci\'on, Desarrollo e Innovaci\'on and Programa Consolider-Ingenio 2010, Spain; the Swiss Funding Agencies (ETH Board, ETH Zurich, PSI, SNF, UniZH, Canton Zurich, and SER); the Ministry of Science and Technology, Taipei; the Thailand Center of Excellence in Physics, the Institute for the Promotion of Teaching Science and Technology of Thailand, Special Task Force for Activating Research and the National Science and Technology Development Agency of Thailand; the Scientific and Technical Research Council of Turkey, and Turkish Atomic Energy Authority; the National Academy of Sciences of Ukraine, and State Fund for Fundamental Researches, Ukraine; the Science and Technology Facilities Council, UK; the US Department of Energy, and the US National Science Foundation.

Individuals have received support from the Marie-Curie programme and the European Research Council and EPLANET (European Union); the Leventis Foundation; the A. P. Sloan Foundation; the Alexander von Humboldt Foundation; the Belgian Federal Science Policy Office; the Fonds pour la Formation \`a la Recherche dans l'Industrie et dans l'Agriculture (FRIA-Belgium); the Agentschap voor Innovatie door Wetenschap en Technologie (IWT-Belgium); the Ministry of Education, Youth and Sports (MEYS) of the Czech Republic; the Council of Science and Industrial Research, India; the HOMING PLUS programme of the Foundation for Polish Science, cofinanced from European Union, Regional Development Fund; the OPUS programme of the National Science Center (Poland); the Compagnia di San Paolo (Torino); the Consorzio per la Fisica (Trieste); MIUR project 20108T4XTM (Italy); the Thalis and Aristeia programmes cofinanced by EU-ESF and the Greek NSRF; the National Priorities Research Program by Qatar National Research Fund; the Rachadapisek Sompot Fund for Postdoctoral Fellowship, Chulalongkorn University (Thailand); and the Welch Foundation, contract C-1845.

\bibliography{auto_generated}

\begin{thebibliography}{10}
\bibitem{Z2}
\hrefCMSnoop {}{}The {\sc pythia6} {\sc Z2} tune is identical to the {\sc Z1}
  tune described in~\cite{Field}, except that {\sc Z2} uses the CTEQ6L PDF,
  while {\sc Z1} uses CTEQ5L.
\end{thebibliography}

\appendix
\clearpage

\cleardoublepage \appendix\section{The CMS Collaboration \label{app:collab}}\begin{sloppypar}\hyphenpenalty=5000\widowpenalty=500\clubpenalty=5000\textbf{Yerevan Physics Institute,  Yerevan,  Armenia}\\*[0pt]
V.~Khachatryan, A.M.~Sirunyan, A.~Tumasyan
\vskip\cmsinstskip
\textbf{Institut f\"{u}r Hochenergiephysik der OeAW,  Wien,  Austria}\\*[0pt]
W.~Adam, E.~Asilar, T.~Bergauer, J.~Brandstetter, E.~Brondolin, M.~Dragicevic, J.~Er\"{o}, M.~Flechl, M.~Friedl, R.~Fr\"{u}hwirth\cmsAuthorMark{1}, V.M.~Ghete, C.~Hartl, N.~H\"{o}rmann, J.~Hrubec, M.~Jeitler\cmsAuthorMark{1}, V.~Kn\"{u}nz, A.~K\"{o}nig, M.~Krammer\cmsAuthorMark{1}, I.~Kr\"{a}tschmer, D.~Liko, T.~Matsushita, I.~Mikulec, D.~Rabady\cmsAuthorMark{2}, B.~Rahbaran, H.~Rohringer, J.~Schieck\cmsAuthorMark{1}, R.~Sch\"{o}fbeck, J.~Strauss, W.~Treberer-Treberspurg, W.~Waltenberger, C.-E.~Wulz\cmsAuthorMark{1}
% ...
\vskip\cmsinstskip
% ...
1:~~Also at Vienna University of Technology, Vienna, Austria\\
2:~~Also at CERN, European Organization for Nuclear Research, Geneva, Switzerland\\
% ...

\end{sloppypar}
\end{document}